\documentclass[a4paper]{jpconf}
\usepackage{xcolor}
\usepackage{graphicx}
\usepackage{caption}
\usepackage{wrapfig}
\usepackage{subcaption, multirow}
\usepackage{relsize}
\usepackage{amsmath}
\usepackage{MnSymbol}
\usepackage{wasysym}
\usepackage[labelfont=bf]{caption}
\begin{document}
\title{Are the dynamics of wall turbulence in minimal channels and larger domain channels equivalent? A graph-theoretic approach}

\author{Ahmed Elnahhas$^1$, Emma Lenz$^2$, Parviz Moin$^1$, Adri\'an Lozano-Dur\'an$^3$ and H. Jane Bae$^2$}

\address{$^1$ Center for Turbulence Research, Stanford University, Stanford CA, 94305 USA}
\address{$^2$ Graduate Aerospace Laboratories, California Institute of Technology, Pasadena CA, 91125 USA}
\address{$^3$ Department of Aeronautics \& Astronautics, Massachusetts Institute of Technology, Cambridge MA, 02139 USA}

\ead{ahmed97@stanford.edu}

\begin{abstract}

This work proposes two algorithmic approaches to extract critical dynamical mechanisms in wall-bounded turbulence with minimum human bias. In both approaches, multiple types of coherent structures are spatiotemporally tracked, resulting in a complex multilayer network. Network motif analysis, i.e., extracting dominant non-random elemental patterns within these networks, is used to identify the most dominant dynamical mechanisms. Both approaches, combined with network motif analysis, are used to answer whether the main dynamical mechanisms of a minimal flow unit (MFU) and a larger unconstrained channel flow, labeled a full channel (FC), at $Re_\tau \approx 180$, are equivalent. The first approach tracks traditional coherent structures defined as low- and high-speed streaks, ejections, and sweeps. It is found that the roll-streak pairing, consistent with the current understanding of self-sustaining processes, is the most significant and simplest dynamical mechanism in both flows. However, the MFU has a timescale for this mechanism that is approximately $2.83$ times slower than that of the FC. In the second approach, we use semi-Lagrangian wavepackets and define coherent structures from their energetic streak, roll, and small-scale phase space. This method also shows similar motifs for both the MFU and FC. It indicates that, on average, the most dominant phase-space motifs are similar between the two flows, with the significant events taking place approximately $2.21$ times slower in the MFU than in the FC. This value is more consistent with the implied timescale ratio of only the slow speed streaks taking part in the roll-streak pairing extracted using the first multi-type spatiotemporal approach, which is approximately $2.17$ slower in the MFU than the FC.

\end{abstract}

\section{Introduction}

It has long been hypothesized that wall-bounded turbulence can be decomposed into a set of coherent structures that coexist and interact \cite{Robinson1991}. This concept can be interpreted kinematically by synthesizing the turbulent flow field as a superposition of building block eddies; the attached eddy model (AEM) adopts this approach for high-$Re_\tau$ flows using a hierarchy of eddies scaling with distance from the wall \cite{Townsend1976,Marusic2019}. A more comprehensive model would account for the dynamics of the building block eddies, both in isolation as well as for their interactions. The most famous dynamic coherent structure model of wall-bounded turbulence is the roll-streak self-sustaining process (SSP) \cite{Jimenez1991,Hamilton1995,Jimenez1999,Flores2010,HwangCossu2010,HwangCossu2011,baectr2019,baejfm2021,lozano2020}. This SSP is primarily extracted from simulations of minimal flow units (MFU) at low $Re_\tau$, which isolate a single scale of wall-bounded turbulence, as well as a single instance of the SSP at a time by enforcing severe periodicity \cite{Jimenez1991,Hamilton1995,Jimenez1999,lozano2020}. These MFU simulations have also been extended to larger scales by isolating structures in the logarithmic region of the flow where similar dynamical cycles were observed \cite{Flores2010,HwangCossu2010,HwangCossu2011,baectr2019,baejfm2021}. Taking the AEM, one can theorize that a ``dynamic"-AEM can be constructed based on the findings from MFUs of the buffer and logarithmic regions of the flow by superposing SSPs at each scale of motion and neglecting strong interscale interactions. However, recent experimental observations of inner/outer modulation and large-scale streak aggregation suggest more to the story \cite{Hutchins2007,Hutchins2007a,Marusic2010,zhou2022}. Hence, a few questions immediately arise concerning the current understanding of wall-bounded turbulence: Are the dynamics extracted from MFUs relevant to wall-bounded turbulent flows in large unbounded domains at the same $Re_\tau$? Or do interactions between different instances of SSPs heavily influence each other, perhaps altering their timescales? Are other, possibly dominant, dynamic cycles suppressed in the artificially restricted domains of MFUs? Finally, what is the importance of interscale interactions of different-sized SSPs at larger $Re_\tau$?  To understand the dynamical interactions amongst the same scales of motion and inter-scale, we need to find a systematic and algorithmic way to extract and study these dynamics. In this study, we propose such a framework and apply it first to study whether the dynamics of the MFU exist unaltered in large domain channel simulations, henceforth referred to as full channels (FC), at $Re_\tau=180$, or not.

Recently, coherent structure identification and temporal tracking algorithms have been developed to understand the basic dynamical properties of select eddy definitions, such as momentum-carrying structures, velocity streaks, and energy-containing eddies \cite{lozano2012,lozano2014time,Dong2020,bae2021}. These algorithms are based on spatial and temporal clustering of grid points that satisfy certain thresholding and proximity criteria. The result is a directed graph where the nodes represent the coherent structures, and the edges represent temporal connections between the structures. Therefore, advanced graph processing tools can extract information about these coherent structures' spatiotemporal properties. Extending the clustering and tracking algorithms to track several definitions of eddies simultaneously, along with the interconnections between eddy types, results in a higher dimensional graph, where the eddy type acts as the new dimension, called a multilayered network~\cite{Boccaletti2014,kivela2014}. We use network motifs and graphlet search algorithms~\cite{Milo2002,Ribeiro2021} to analyze and compare dynamically significant patterns extracted from the buffer layer MFU and large domain channel flows.

\begin{figure}
    \centering
    \begin{subfigure}[c]{0.95\textwidth}
        \includegraphics[trim = {0cm 3cm 0cm 0cm},width=\textwidth]{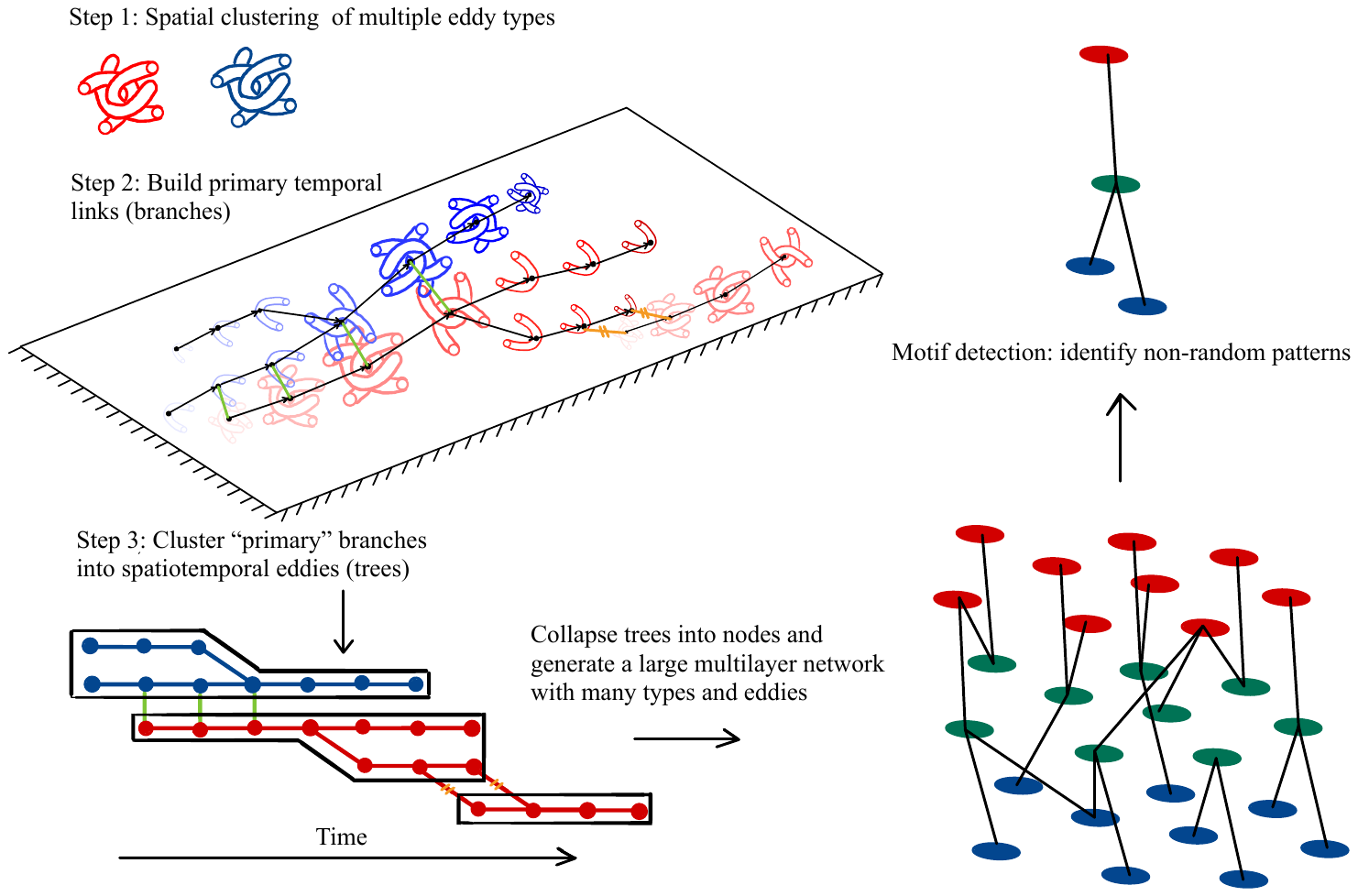}
        \caption{}
        \label{fig:MotifDetectionRepresentation}
    \end{subfigure}
    \begin{subfigure}[c]{0.75\textwidth}
        \includegraphics[width=\textwidth]{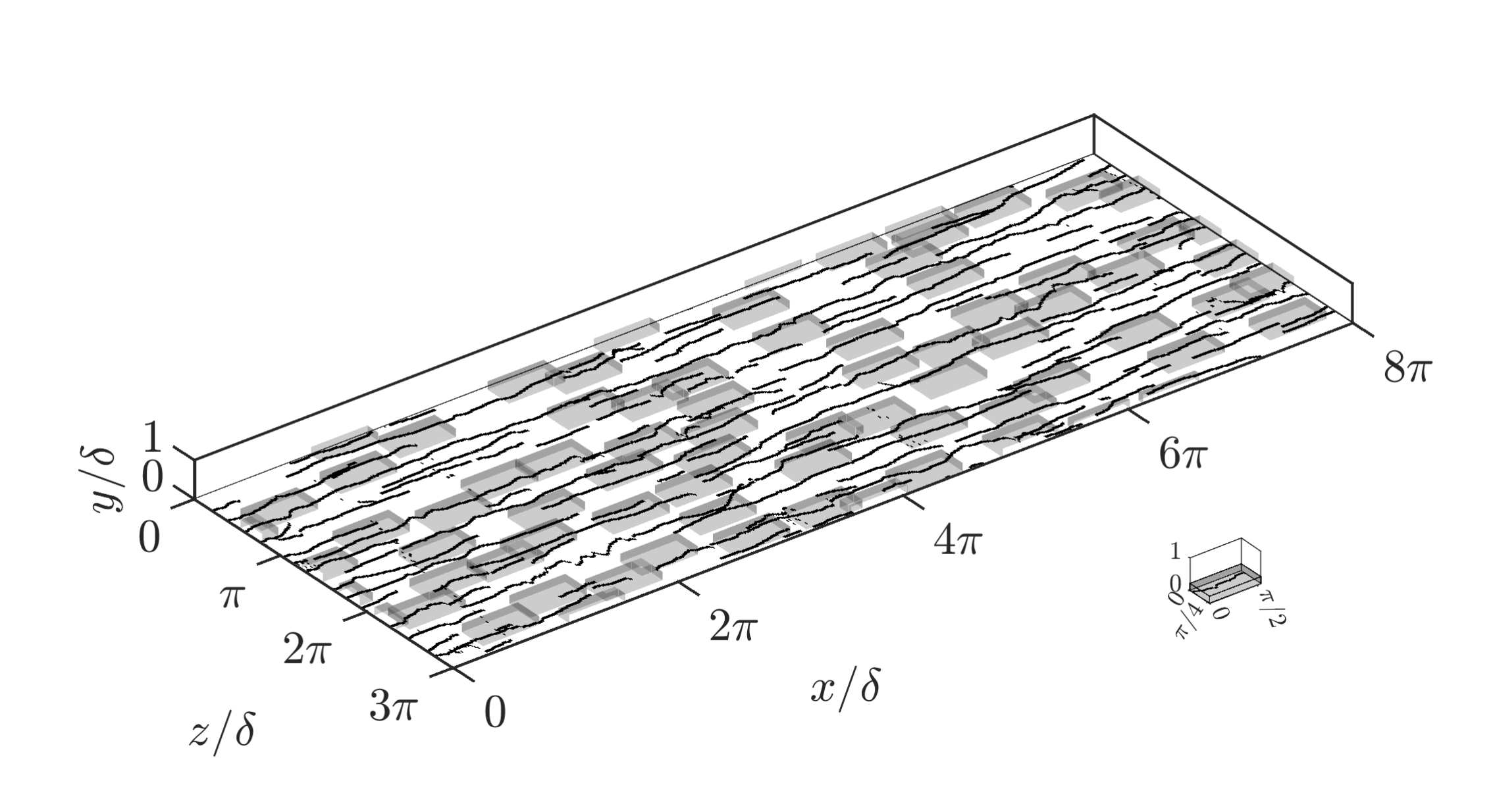}
        \caption{}
        \label{fig:WavePacketRepresentation}
    \end{subfigure}
    \caption{(a) The proposed multi-type spatiotemporal tracking and motif detection methodology. Different types of eddies are clustered simultaneously, and the resulting multilayer network is searched for patterns called network motifs. The lighter structures represent earlier times in the evolution of a spatiotemporal eddy. Green lines represent the intertype connections transferred from single-time connections to branches to trees. Orange cross-lines indicate intra-type branch links that are not considered simultaneously ``primary" for tree clustering. (b) The proposed semi-Lagrangian wavepacket approach where the greyed areas are MFU-sized boxes, which are allowed to advect and collect quantities for the phase-space-temporal dynamical analysis. Black lines are the spanwise centroids of low-speed streaks projected onto the wall. The MFU is shown for comparison.}
    \label{fig:SchematicIdea}
\end{figure}

A network motif of a complex network is a pattern occurring at a significantly higher frequency than in a random network of the same size \cite{Milo2002}. Network motifs have been identified in various biological networks, such as those responsible for gene expression and neuronal cell interactions \cite{Alon2007}. They have also been identified in larger-scale ecological and engineering networks \cite{Milo2002}. Since our directed multilayer network has time embedded in its construction, detecting motifs accounts for the most significant dynamical patterns between eddies of various types directly. Figure \ref{fig:SchematicIdea}(a) illustrates the proposed approach. Each node in the network represents the full spatiotemporal history of a coherent structure, and the edges represent the interactions between two coherent structures at some point during their evolution. In the context of the current question, finding the most dominant interaction pattern compared to a set of random networks helps us isolate the key dynamical mechanism in both the FC and the MFU and compare their similarity. Applications of network analysis to fluid flow are only recent phenomena~\cite{lozano2014time,Taira2016,schmid2018,Iacobello2019,Krueger2019,Fernex2021,Iacobello2021,Li2021,Perrone2021,Taira2022}, and applications to the study of the dynamics of coherent structures have not been attempted in fully developed wall-bounded turbulent flows. The results of this spatiotemporal approach are supplemented with motifs extracted from an energetic phase-space constructed using semi-Lagrangian wavepackets illustrated in figure \ref{fig:SchematicIdea}(b).

The rest of the paper is organized as follows. Section \ref{Methods} describes the datasets utilized to perform the analysis. In section \ref{CST}, the multi-type coherent structure tracking and motif detection algorithms are described, and the results of their application to the MFU and the FC at $Re_\tau=180$ are discussed. In section \ref{WPF}, the semi-Lagrangian wavepackets and phase-space temporal motif analysis approach is explained, and the results of applying it to the same datasets are discussed. Finally, conclusions and future work are presented in section \ref{Conclusions}.

\section{Simulation details}\label{Methods}

Let $u_\tau$ and $\nu$ denote the flow's friction velocity and kinematic viscosity, respectively. The two flows analyzed are direct numerical simulations (DNS) of an MFU and an FC inside a channel of half-height $\delta$. The friction Reynolds number for both flows is $Re_\tau \equiv \delta u_\tau/\nu \approx 180$. In this flow, $x$, $y$, and $z$ are the streamwise, wall-normal, and spanwise directions, respectively, corresponding to their respective mean and fluctuating velocities $U$ and $u$, $V$ and $v$, and $W$ and $w$. The domain of the MFU has a size of $\{0.564\pi\times2\times0.282\pi\}\delta$, and that of the FC has a size of $\{8\pi\times2\times3\pi\}\delta$. The resolution in inner units of both simulations are $\{\Delta x^+, \Delta y^+_{min},\Delta y^+_{max}, \Delta z^+\} = \{6.05, 0.17, 7.52, 6.05\}$, which is consistent with prior DNS simulations. Flow fields are stored every $\Delta t^+ \approx 1$ for $T^+ = 2000$ and $T^+ = 25000$ inner time units for the FC and MFU, respectively. These correspond to approximately $11\delta/u_\tau$ and $135\delta/u_\tau$, respectively.

\section{Multi-type spatiotemporal coherent-structure tracking and motif detection}\label{CST} 
\subsection{Multilayer network generation}
The multi-type coherent-structure tracking algorithm is a three-step process, schematically illustrated in figure \ref{fig:SchematicIdea}(a): 

Step 1: In this work, the eddies considered are the low- and high-speed streaks and wall-normal streamwise momentum flux eddies, namely ejections and sweeps, interchangeably referred to as Q2 and Q4, respectively~\cite{lozano2012,bae2021}. At each time instant, each of these eddies is defined as the set of connected points in space, $\Omega$, satisfying the following inequalities

\begin{equation}\label{MomDef}
    \big\{(x,y,z):|uv| > \alpha_{uv} u'(y)v'(y) \quad \& \quad
    \begin{cases}
    \bigg(\frac{\int_\Omega u~dV_{\Omega}}{\int_\Omega dV_{\Omega}} < 0\bigg)~\&~\bigg(\frac{\int_\Omega v~dV_{\Omega}}{\int_\Omega dV_{\Omega}} > 0\bigg)\bigg\} \implies \mathrm{Ejections~(Q2)}\\[10pt]
    \bigg(\frac{\int_\Omega u~dV_{\Omega}}{\int_\Omega dV_{\Omega}} > 0\bigg)~\&~\bigg(\frac{\int_\Omega v~dV_{\Omega}}{\int_\Omega dV_{\Omega}} < 0\bigg)\bigg\} \implies \mathrm{Sweeps~(Q4)}
    \end{cases}\mathrm{,}
\end{equation}
\begin{equation}\label{StreakDef}
    \big\{(x,y,z) : \sqrt{u^2+w^2} > \alpha_{str}u'(y) \quad \& \quad
    \begin{cases}
    u<0\big\} \implies \mathrm{Slow~Streak} \\
    u>0\big\} \implies \mathrm{Fast~Streak}
    \end{cases}{,}
\end{equation}
where $u'(y)$ and $v'(y)$ are the root-mean-squared streamwise and wall-normal velocity fluctuations extracted from the corresponding simulation, i.e., the MFU and the FC, $\alpha_{\{uv,str\}}$ are scaling coefficients for the thresholds, and $V_{\Omega}$ is the physical volume associated with the set of points, $\Omega$. The clustering process is known to have a percolation crisis, and we use the values $\{\alpha_\tau,\alpha_{str}\}=\{1.30,1.75\}$, which maximize the number of clusters identified per timestep on average. Each region, $\Omega$, is labeled as a coherent structure at some instant in its evolution. Once subdivided into these four sets, contiguous time snapshots construct their temporal evolution.
\begin{figure}
    \centering
    \begin{subfigure}[c]{0.40\textwidth}
        \includegraphics[width=\textwidth]{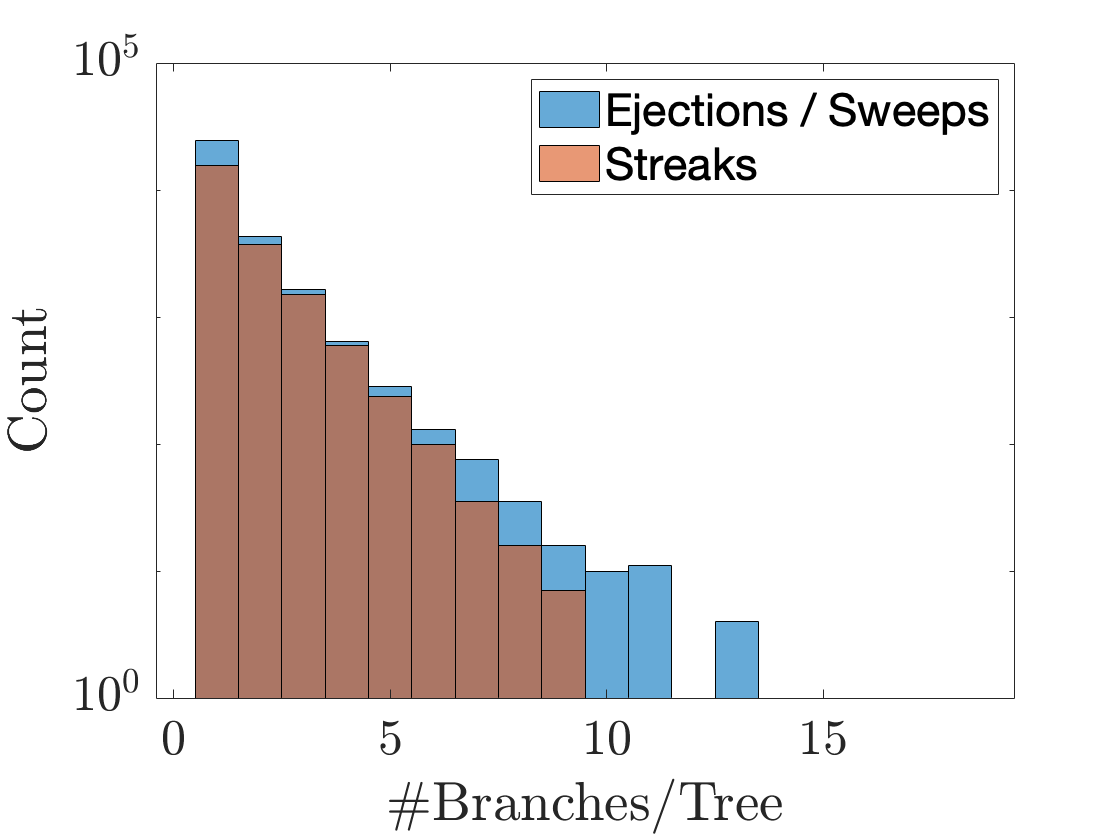}
        \caption{}
        \label{fig:BranchToGraphGrouping}
    \end{subfigure}
    \begin{subfigure}[c]{0.58\textwidth}
        \includegraphics[width=\textwidth]{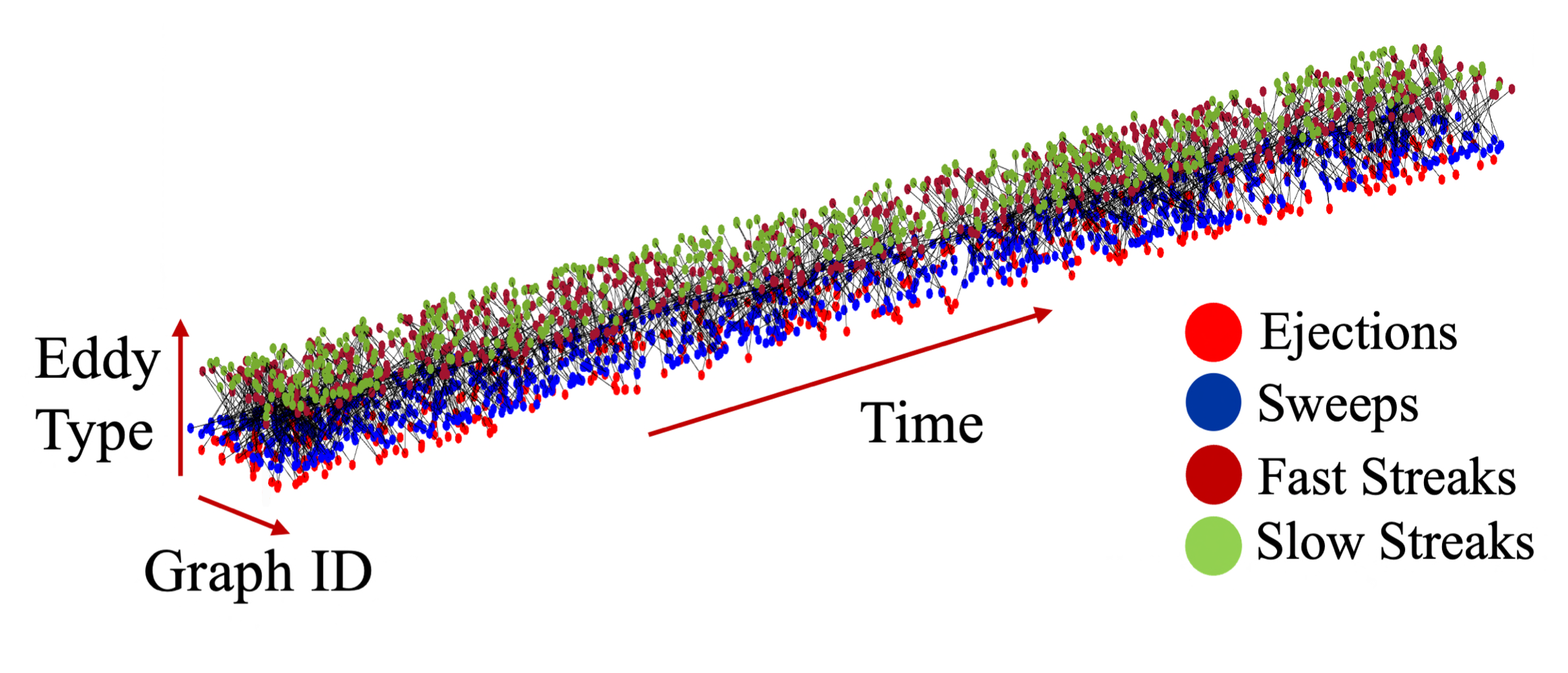}
        \caption{}
        \label{fig:MultitypeGraph}
    \end{subfigure}
    \caption{(a) Histogram indicating the distribution of how many individual attached eddy branches are grouped into trees using the simultaneous primary branch interaction metric in Equation \eqref{PrimaryMetric}. (b) A multilayer network with each node representing the full spatiotemporal evolution of an attached eddy tree placed at its temporal centroid. The links between the nodes indicate an interaction or overlap at some point during the lifetime of the two linked spatiotemporal eddies. Only a subset of the complete network in the FC is shown.}
    \label{fig:TrackingClusters&Graphs}
\end{figure}

Step 2: The details of the structure tracking in time are described in \cite{lozano2014time}. In summary, object volumetric overlap between timesteps is used to construct a graph with edges representing connections in time and objects at different time steps representing nodes. While previously this was done for only a single type of object, this procedure is repeated $N^2$ times where $N$ is the number of eddy types. The graph is first clustered into branches, which are the links between objects that mutually experience the least relative change from one timestep to the next. 

Step 3: The branches are then further partitioned to generate subgraphs or trees, which are collections of branches that interact in a ``primary" way. We define a primary branch connection using the relative overlap metric from branch $i$ to $j$ of the same eddy type and vice versa, which is defined as follows
\begin{equation}\label{PrimaryMetric}
    O_{i\rightarrow j} = \frac{\int V_i \cap V_j dt_i}{\int V_i dt_i} \quad \text{and} \quad  O_{j\rightarrow i} = \frac{\int V_j \cap V_i dt_j}{\int V_j dt_j} \text{,}
\end{equation}
where the integration is over the lifetime of the source branch and $V_i$ is the volume of a branch at a given time across its lifetime. The intersecting volume is computed at the same absolute time for both branches. The branch link $j$ to branch $i$ with the highest overlap metric is considered the primary branch in the positive or negative time directions. If two branches are each other's primary links, they are clustered into a subgraph or tree. This process is applied recursively, and each resulting tree is subsequently referred to as a spatiotemporal eddy. An illustration of the partitioning when two branches do not point to each other as primary branch connections is represented with the crossed orange lines in figure \ref{fig:SchematicIdea}(a).

During step 2 and step 3 of the process, the physical connections between eddies of different types are transferred from single-time connections at the end of step 1 to connections between branches and, finally, trees at the end of steps 2 and 3, respectively. This results in a large, complex, multilayered network that encodes spatiotemporal interactions between eddies of the same type and those between eddies of different types. These intertype connections are represented with green lines in the schematic illustration of figure \ref{fig:SchematicIdea}(a).

After the three-step process, many of the spatiotemporal eddies are small structures existing in the center of the channel, which, if tracked, result in an overly large/complex multilayer network. To ease the subsequent analysis, we limit ourselves to interactions between attached eddies that live for a minimal amount of time, i.e., those who, at some point during their lifetime, had a minimal wall-normal height (on either wall) of $y^+_{min} < 20$ \cite{lozano2014time,bae2021}, and have lived for longer than $T^+_{life} = 20$. Furthermore, they must have interacted with another attached eddy during their lifetime. The volumes occupied by the ejections, sweeps, slow, and fast streaks that satisfy these criteria and subsets of them in both the MFU and the FC are presented in table \ref{tab:VolumeEddy}. The applied criteria only restrict the minority of the attached eddies by volume, approximately $11.5-17.5\%$, from participating in the subsequent motif analysis. Figure \ref{fig:TrackingClusters&Graphs}(a) showcases the result of the step 3 partitioning of branches into trees using the overlap metric for attached eddies, where the size of the trees in terms of branches is exponentially decaying, and no single tree contains more than $\mathcal{O}(10)$ branches. Given this partitioning, figure \ref{fig:TrackingClusters&Graphs}(b) shows a subset of the multilayer network that emerges due to all the interactions of attached spatiotemporal ejections, sweeps, and slow and fast streaks in the FC where each dot contains within it the entirety of the detailed spatiotemporal evolution of each of the spatiotemporal eddies.  

\begin{table}
\caption{\label{tab:VolumeEddy} The average percentage of the domain's volume occupied by certain types of eddies that satisfy particular criteria in both the full channel and minimal flow unit. $\mathcal{A}:{y_{min}^+<20}$. $\mathcal{T}:{T_{life}^+>20}$. $\mathcal{I}:\text{Interaction with another $\mathcal{A}-$\text{valid eddy}}$. $\mathcal{S}_1^{\{F,M\}}:\text{Involved in most significant roll-streak motif}$. $\mathcal{F}_2^M:\text{Involved in similar motif only presented in MFU}$. The two values for $\mathcal{S}_1^M$ and $\mathcal{F}_2^M$ eddies are averages performed over the entire temporal extent of the simulation and only over the times when constituent eddies in the motifs are alive, respectively.}
\begin{center}
\renewcommand{\arraystretch}{1} 
\begin{tabular}{ccccccc}
 \br
 \multicolumn{3}{c}{Minimal Flow Unit}                                                                      && \multicolumn{3}{c}{Full Channel}          \\
 \mr
  Filter & Q2 \& Q4 & Streaks && Filter & Q2 \& Q4 &  Streaks \\
  \mr  
\small$\mathcal{A}$&5.12\%&2.55\%&&\small$\mathcal{A}$&5.19\%&2.43\%\\
\small$\mathcal{A}~\&~\mathcal{T}$&4.90\%&2.52\%&&\small$\mathcal{A}~\&~\mathcal{T}$&4.52\%&2.15\%\\
\small$\mathcal{A}~\&~\mathcal{T}~\&~\mathcal{I}$&4.43\%&2.31\%&&\small$\mathcal{A}~\&~\mathcal{T}~\&~\mathcal{I}$&4.16\%&2.10\%\\
\small$\mathcal{A}~\&~\mathcal{T}~\&~\mathcal{I}~\&~\mathcal{S}^M_1$&1.09-3.87\%&0.72-2.44\%&&\small$\mathcal{A}~\&~\mathcal{T}~\&~\mathcal{I}~\&~\mathcal{S}^F_1$&2.04\%&1.01\%\\
\small$\mathcal{A}~\&~\mathcal{T}~\&~\mathcal{I}~\&~\mathcal{F}^M_2$&2.56-4.14\%&1.10-2.15\%&&|&|&|\\
 \br
\end{tabular}
\end{center}
\end{table}

\subsection{Motif detection}
The resulting multilayer network, at the tree level, is fed into an open-source motif detection code, FAst Network MOtif Detection (FANMOD) \cite{wernicke2006fanmod}, which uses the exhaustive enumerate subgraph (ESU) algorithm to enumerate all isomorphic subgraphs, $s$, of a certain size in our multilayer network \cite{wernicke2006ESU}. The frequency of each isomorphic subgraph $s$, given by $m(s)$, is compared with statistics from a set of randomly generated networks to determine over-representation in the original network, i.e., significant motifs. 

In this case, the randomly generated networks are constructed from our original network by randomly switching edges between nodes in the network while maintaining its size, $|G|$. The sizes of the original network for each flow are $|G|_{FC} = 31855$, and $|G|_{MFU}=1251$. Since their criteria are mutually exclusive, ejections and sweeps cannot form connections, as is the case between fast and slow streaks. This constraint is reflected in the random networks generated by maintaining the local colored degree vector of each node during the randomization \cite{Ribeiro2014}. This means that, for example, if a sweep were originally connected to two fast streaks and one slow streak, it would remain connected to two fast streaks and one slow streak after the randomization. We can, therefore, interpret the randomization as a reordering in space and time of the spatiotemporal eddies while maintaining their allowable interactions. 

The frequency of each subgraph in the random graph is given by $m_r(s)$. The mean and variance of $m_r(s)$ are computed from an ensemble of random graphs and denoted $\overline{m_r(s)}$ and $Var(m_r(s))$, respectively. Enough random graphs were generated so that the mean and the variance converged. This was done with $100$ and $1000$ random graphs for the FC and MFU, respectively. A significance score was determined for each isomorphic subgraph based on the random mean and standard deviation. Furthermore, to account for shifts in the scores associated with the differences in the sizes of the networks, a normalization based on the law of large numbers utilizing $|G|$ is added. The significance score is, therefore,
\begin{equation}\label{SignificanceScore}
    \widetilde{W}(s) = \frac{m(s) - \overline{m_r(s)}}{\left(Var(m_r(s))|G|\right)^{1/2}}.
\end{equation}

\subsection{Results}

The results of the motif analysis indicate that there are no $3$-node motifs, i.e., none where significantly different than random,  and we focus our discussions on $4$-node motifs. The schematics in table \ref{tab:SpatioTempMotifZFreq} illustrate the results for the $4$-node motifs. The motifs are labeled as  $\{\mathcal{S},\mathcal{F}\}^{\{F,M\}}_\#$ with $\mathcal{S}$ or $\mathcal{F}$ denoting motifs ordered by significance score and frequency, respectively, and the superscripts $F$ and $M$ denoting the flow, FC and MFU, respectively. The left portion of the schematics in the table shows the top six to seven motifs in descending order of significance score. The two flows share the most significant motifs, where the top six in the MFU are found in the top seven of the FC. However, besides the most dominant motif, the order of the other five is shifted around in the MFU with respect to the FC, as denoted by the two-headed black arrows. The most dominant $4$-node motif for both flows is a pairing of a slow and a fast streak interacting with an ejection and a sweep, consistent with the roll-streak SSP, labeled from now on as $\mathcal{S}_1^{\{F,M\}}$. This likely indicates that the roll-streak pairing is, in fact, the simplest form of spatiotemporal organization exhibited in these single-scale wall-bounded flows. The right portion of the schematics in the table shows the top six motifs in descending order of frequency. The two flows share five of the top six between them. However, their orders are shifted around in a similar manner to the top six most dominant motifs. Furthermore, motif $\mathcal{F}_2^M$ strongly resembles the structure of $\mathcal{S}_1^{\{F,M\}}$ except for a missing link between slow streaks and sweeps. This is likely due to the confined domain restricting the emergence of the structures simultaneously. Whether this confinement leads to some ordering in the emergence and death of the constituent eddies of this motif is examined further below.

Furthermore, table \ref{tab:SpatioTempMotifZFreq} reports the significance scores and frequency of occurrence of each motif. Note that while the frequency of the most significant motif in both flows seems to be low compared to all possible $4$-node induced subgraphs in either $G_{FC}$ or $G_{MFU}$, that does not mean that the constituent eddies are few or not substantial. In fact, table \ref{tab:VolumeEddy} shows that approximately $50\%$ of the attached eddies by volume in the FC participate in this motif. For the MFU, the volume occupied by these eddies averaged over the entire temporal extent of the simulation indicates that only $25-30\%$ of them participate in the motif. However, there are times when no eddies participating in this motif are present in the simulation domain. Thus, if we condition the average on the times when the eddies participating in the motif are present, i.e., alive, the average volume of ejections and sweeps is around $87\%$ of their unconditioned average volume, and that of streaks is even larger than the unconditioned average volume of attached streaks throughout the simulation. A plausible explanation is as follows: In the MFU, on average, only a single instance of an SSP can exist at a time, and once the large-scale streaky structures break down in what is known as bursting \cite{jimenez2005bursting}, the current tracking procedure sees no attached structures. While this also happens in the case of the FC,  since there is always at least one instance of this motif active in the domain, conditionally averaging over the times when a constituent eddy is alive is redundant. However, for the MFU, there are time periods where we are averaging over zero volume due to the presence of only small scales. Overall, this implies that in subsequent analyses, we should include the simultaneous tracking of small scales in the motifs we search for to find the full cycle of the SSP. Finally, it is interesting to note that all the most frequent yet marginally significant motifs, except for $\mathcal{F}_2^M$, in both the FC and MFU have an equal significance score in each flow, respectively. This indicates that their frequencies are linked during the construction of the random networks and that they are most likely kinematically constrained to one another. We, therefore, focus our dynamical analyses on the eddies participating in each of the motif instances of $\mathcal{S}_1^{\{F,M\}}$, albeit including data from motif $\mathcal{F}^M_2$ in the case of the MFU due to its similarity to $\mathcal{S}_1^M$.

\definecolor{Ejection}{RGB}{254,1,0}
\definecolor{Sweep}{RGB}{4,70,143}
\definecolor{Faststreak}{RGB}{117,0,0}
\definecolor{Slowstreak}{RGB}{127,197,55}

\begin{table}
\caption{\label{tab:SpatioTempMotifZFreq}The dynamical $4$-node motifs detected using the ESU algorithm implemented in FANMOD in both the FC and the MFU using the aforementioned random network model. The left portion shows motifs ordered in descending order of significance score, indicating the strongest non-random events in both domains. The right portion shows motifs ordered in descending order of frequency. The reported motifs exclude any with a significance score below $\widetilde{W}(s)<0.05$. The top and bottom rows of motif schematics represent those extracted from the FC and the MFU, respectively. Above and below the motif schematics, the significance score, $\widetilde{W}(s)$, and frequency percentage, $\mathcal{F}(s)$, i.e., the percentage of occurrences in the original networks with respect to the total number of all 4-node subgraphs present in the MFU and FC networks, which were 1811927 and 7057, respectively, are presented along with their labels. The bounding red box indicates that the most non-random dynamical pattern, $\mathcal{S}_1^{\{F,M\}}$, is the roll-streak pairing and is consistent amongst both flow domains. The bounding dashed purple box shows a similar motif to $\mathcal{S}_1^{\{F,M\}}$, $\mathcal{F}_2^{M}$, that only emerges in the MFU. \textcolor{Ejection}{$\newmoon$} $\rightarrow$ Ejections. \textcolor{Sweep}{$\newmoon$} $\rightarrow$ Sweeps. \textcolor{Faststreak}{$\newmoon$} $\rightarrow$ Fast streaks. \textcolor{Slowstreak}{$\newmoon$} $\rightarrow$ Slow streaks.}
\begin{center}
\renewcommand{\arraystretch}{1} 
\begin{tabular}{c|ccccccc|cccccc}
\mr
&\multicolumn{7}{c}{Significance Ordering}&\multicolumn{6}{c}{Frequency Ordering}\\
\mr
FC&$\mathcal{S}_1^F$&$\mathcal{S}_2^F$&$\mathcal{S}_3^F$&$\mathcal{S}_4^F$&$\mathcal{S}_5^F$&$\mathcal{S}_6^F$&$\mathcal{S}_7^F$&$\mathcal{F}_1^F$&$\mathcal{F}_2^F$&$\mathcal{F}_3^F$&$\mathcal{F}_4^F$&$\mathcal{F}_5^F$&$\mathcal{F}_6^F$\\
\mr
$\widetilde{W}(s)$&4.71&3.76&3.50&3.47&3.13&3.04&2.56&0.13&0.13&0.13&0.13&0.13&0.13\\
$\mathcal{F}(s)$&0.26&0.19&0.16&0.14&0.15&0.09&0.09&7.28&7.10&6.10&4.95&3.53&3.36\\
\mr
&\multicolumn{13}{c}{\includegraphics[width=0.915\textwidth,trim={5cm 5.5cm 0.25cm 11cm},clip]{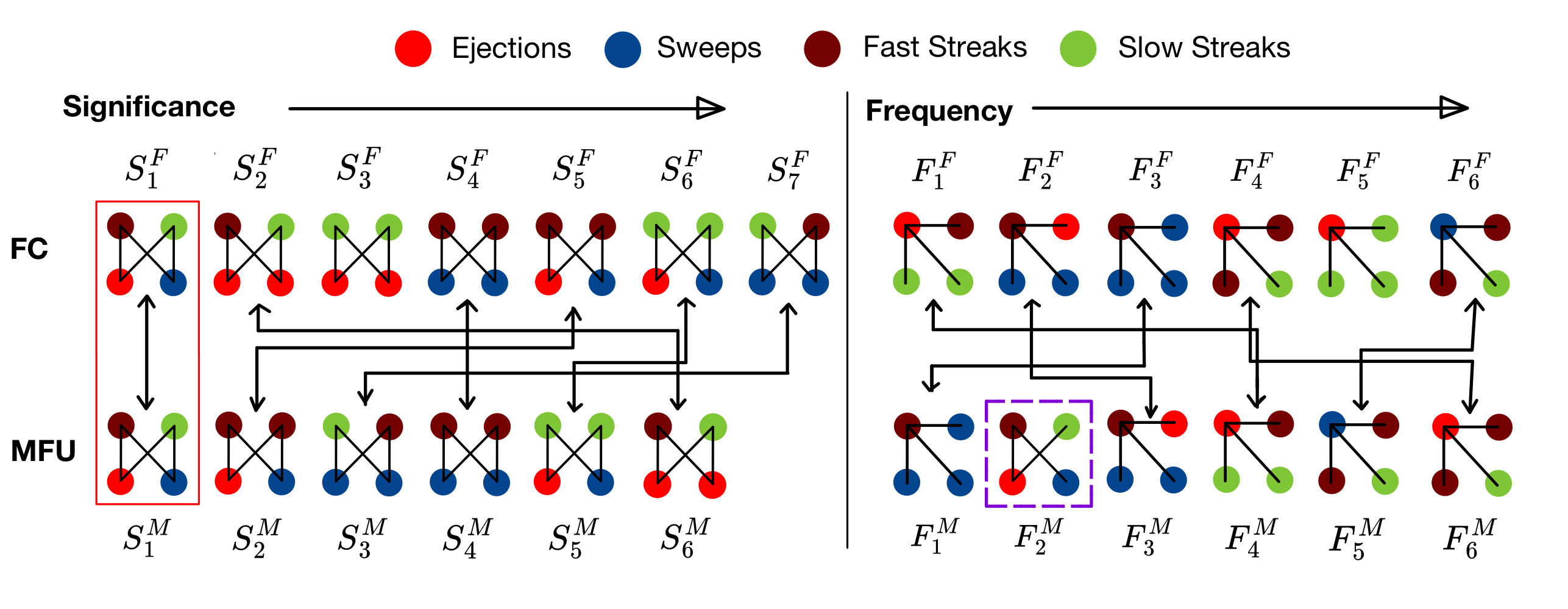}}\\
\mr
MFU&$\mathcal{S}_1^M$&$\mathcal{S}_2^M$&$\mathcal{S}_3^M$&$\mathcal{S}_4^M$&$\mathcal{S}_5^M$&$\mathcal{S}_6^M$&|&$\mathcal{F}_1^M$&$\mathcal{F}_2^M$&$\mathcal{F}_3^M$&$\mathcal{F}_4^M$&$\mathcal{F}_5^M$&$\mathcal{F}_6^M$\\
\mr
$\widetilde{W}(s)$&1.54&1.30&0.83&0.79&0.73&0.38&|&0.37&0.05&0.37&0.37&0.37&0.37\\
$\mathcal{F}(s)$&0.72&0.47&0.34&0.35&0.14&0.10&|&15.20&8.86&8.67&3.83&3.49&3.25\\
\mr
\end{tabular}
\end{center}
\end{table}

\begin{figure}
    \centering
    \includegraphics[width=0.5\textwidth]{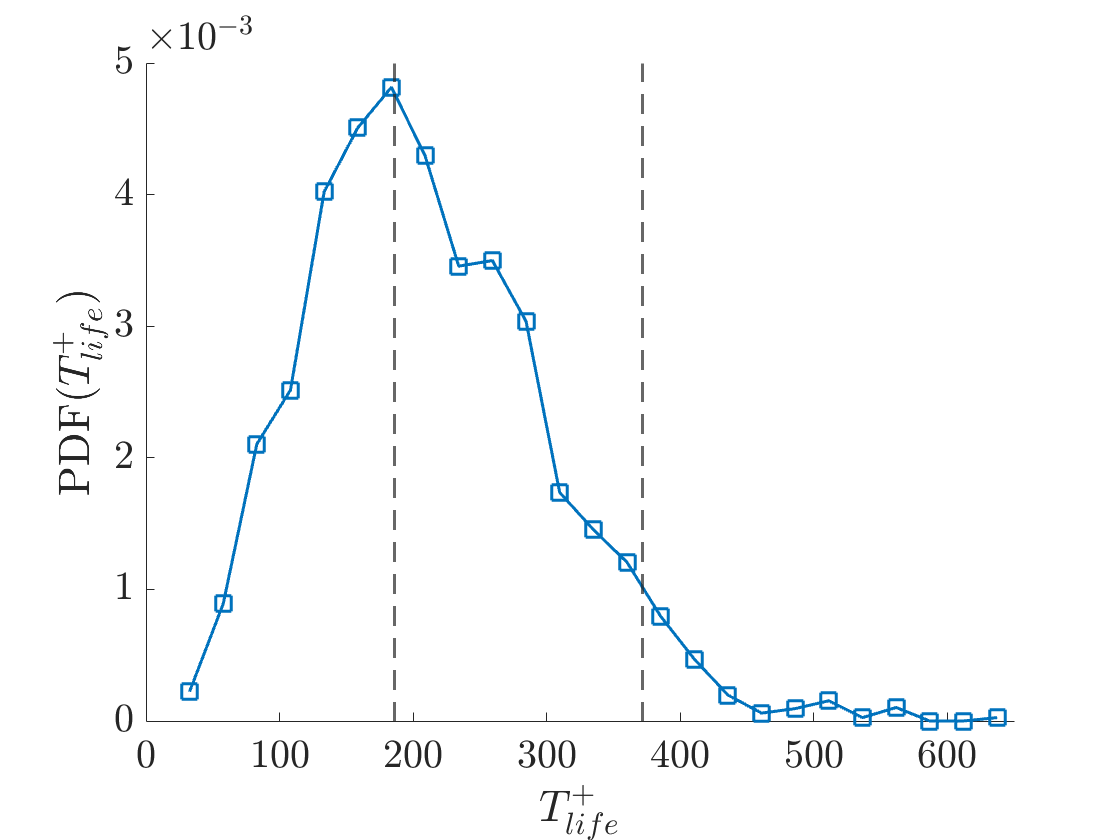}
    \caption{PDF of the lifetime of the roll-streak motif instances, $\mathcal{S}_1^F$, extracted from the FC. The two vertical dashed lines are placed at $\delta/u_\tau$ and $2\delta/u_\tau$, respectively.}
    \label{fig:RollStreakLifeTimeFC}
\end{figure}

Table \ref{tab:AdvectionVelocityLifeTime} shows the average lifetimes of each motif, $\overline{T^+}_{life}$, defined as the time from when the first participating eddy emerges to when the final participating eddy dies, in both the FC and MFU for motifs $\mathcal{S}_1^{\{F,M\}}$ and $\mathcal{F}_2^M$. Simultaneously, it shows the average advection velocity, $\overline{u^+}_{adv}$, of the participating eddies in the streamwise direction. It is observed that $\overline{T^+}_{life}$ is approximately $2.83$ times longer in the MFU compared to its FC counterpart. The advection velocities of the constituent eddies of these motifs, $\overline{u^+}_{adv}$, range from $1.3$ to $3.1$ times slower in the MFU compared to its FC counterpart, with the primary reduction present in the ejections and the slow streaks associated with them. This indicates that while the main dynamical mechanism in both flows is the same, they differ in timescales substantially. In fact, it will be shown that the reduction in the speed of slow streaks in the MFU of approximately $2.17$ times compared to those in the FC, is consistent with phase-space motifs extracted using the semi-Lagrangian wavepacket basis analysis to come, which employs the presence of slow streaks as a conditioning variable to determine the advection velocity of the wavepackets. Figure \ref{fig:RollStreakLifeTimeFC} shows the probability density function (PDF) of $T^+_{life}$ for the $\mathcal{S}_1^F$ motif, highlighting that the most probable instances of this motif happen over a single eddy turnover timescale, $\delta/u_\tau$, with a majority ($>95\%$) not lasting longer than two eddy turnover times. 

\begin{table}
\caption{\label{tab:AdvectionVelocityLifeTime}The average lifetime in inner units, $\overline{T^+}_{life}$, of the instances of the most significant motifs in each of the two flows, $\mathcal{S}_1^{\{F,M\}}$, along with the average advection velocity in inner units, $\overline{u^+}_{adv}$, of the constituent eddies taking part in these significant motifs. In the case of the MFU, the eddies taking part in $\mathcal{F}_2^M$ are also included in these averages due to the similarity of the motif and the lack of samples.}
\begin{center}
\renewcommand{\arraystretch}{1} 
\begin{tabular}{ccc}
\mr
\multicolumn{3}{c}{$\overline{T^+}_{life}$}\\
\mr 
Motif Identifier&Minimal Flow Unit&Full Channel\\
\mr
$\mathcal{S}_1^{F,M}$&601&212\\
\mr
\multicolumn{3}{c}{$\overline{u^+}_{adv}$}\\
\mr
Eddy Type&Minimal Flow Unit&Full Channel\\
\mr
Ejections&4.50&14.07\\
Sweeps&11.01&14.30\\
Fast Streaks&9.28&11.86\\
Slow Streaks&6.00&13.00\\
\mr
\end{tabular}
\end{center}
\end{table}

Furthermore, we note that the motifs presented in table \ref{tab:SpatioTempMotifZFreq}, and in particular $\mathcal{S}_1^{\{F,M\}}$, are not partitioned into their $2\times4!$ possible emergence/death orderings. That is because no sufficiently prominent ordering was found in the FC. The emergence ordering that occurred the most, $9.91\%$ of the time, was fast streak $\rightarrow$ sweep $\rightarrow$ slow streak $\rightarrow$ ejection. The death order that occurred the most, $8.53\%$ of the time, was ejection $\rightarrow$ sweep $\rightarrow$ slow streak $\rightarrow$ fast streak. Therefore, we comment on the ordering trends for each individual eddy type, regardless of which permutation this ordering belongs to. Table \ref{tab:EmergenceDeathOrdering} reports these statistics. Overall, in the FC, ejections are more likely to emerge last and die first, sweeps are more likely to emerge at any order and die at any order, albeit with a slightly higher tendency to die first similar to the ejections. On the other hand, fast streaks are more likely to emerge first and die last, whereas slow streaks are slightly more likely to emerge and die in the middle ordering of the motif. It is apparent that the average emergence and death orderings for the fast streaks and the ejections are inverted with respect to each other.

In the MFU, the emergence and death orderings of the ejections are inverted with respect to those of the FC, where they now appear first and die last. This indicates their longer timescales, consistent with their slower advection velocities. Sweeps, however, tend to appear last and die first. Fast streaks are approximately equally likely to appear first, second, or third and die second or third. Finally, slow streaks appear last, with the sweeps, and die either first or third. These differences between the two flows are probably due to the confinement effects of periodic boundary conditions, which the MFU is subject to. 

Finally, streaks, being larger coherent structures than ejections and sweeps, are associated with multiple ejections and sweeps events during their lifetime. This was verified spatiotemporally by examining their type-dependent degree distribution, which is not shown here for brevity. As such, it is possible that a clearer picture of order would appear if, rather than individual motif instances, clusters of individual motif instances are examined \cite{benson2016}.

\begin{table}
\caption{\label{tab:EmergenceDeathOrdering}The percentage of time that ejections, sweeps, fast streaks, and slow streaks emerge and die in a particular order for motifs $\mathcal{S}_1^F$ for the FC, and $\mathcal{S}_1^M$ and $\mathcal{F}_2^M$ for the MFU.}
\begin{center}
\renewcommand{\arraystretch}{1} 
\begin{tabular}{c|cccc|cccc}
\mr
FC&\multicolumn{4}{c}{Emergence(\%)}&\multicolumn{4}{c}{Death(\%)}\\
\mr
Eddy Type&$1^{\mathrm{st}}$&$2^{\mathrm{nd}}$&$3^{\mathrm{rd}}$&$4^{\mathrm{th}}$&$1^{\mathrm{st}}$&$2^{\mathrm{nd}}$&$3^{\mathrm{rd}}$&$4^{\mathrm{th}}$\\
\mr
Ejections&14.2&20.7&24.5&40.6&35.1&27.2&21.8&15.9\\
Sweeps&28.8&25.3&21.6&24.3&32.8&23.2&21.9&22.1\\
Fast Streaks&37.6&26.1&22.1&14.2&11.4&20.1&28.8&39.7\\
Slow Streaks&19.4&31.8&27.9&20.9&20.7&29.5&27.5&22.3\\
\mr
MFU&\multicolumn{4}{c}{Emergence(\%)}&\multicolumn{4}{c}{Death(\%)}\\
\mr
Eddy Type&$1^{\mathrm{st}}$&$2^{\mathrm{nd}}$&$3^{\mathrm{rd}}$&$4^{\mathrm{th}}$&$1^{\mathrm{st}}$&$2^{\mathrm{nd}}$&$3^{\mathrm{rd}}$&$4^{\mathrm{th}}$\\
\mr
Ejections&49.3&24.7&17.9&8.1&9.6&15.8&17.9&56.7\\
Sweeps&13.0&18.9&29.2&38.9&42.9&27.5&19.4&10.2\\
Fast Streaks&25.3&29.6&35.2&9.9&15.2&38.3&28.3&18.2\\
Slow Streaks&12.4&26.8&17.8&43.0&32.3&18.3&34.5&14.9\\
\mr
\end{tabular}
\end{center}
\end{table}

\section{Semi-Lagrangian wavepackets and phase-space temporal motifs}\label{WPF}
\subsection{Defining the size of the semi-Lagrangian wavepackets}
In addition to coherent structures, integral quantities of the flow also provide insight into the flow dynamics. While we can collect statistics on integral quantities directly in the MFU, we require an equally sized domain of integration in the FC to make meaningful comparisons between the two flows. To do so, we utilize semi-Lagrangian wavepackets in the FC to collect statistics on these integral quantities. These wavepackets will be advected downstream following an advection speed to track the relevant structures. To minimize the effect of shearing, we choose the domain to have $y^+ \le 40$, as the structures in this region are known to advect at a similar advection velocity \cite{lozano2014time}. In the wall-parallel directions, the wavepackets, and hence the domain of integration, are sized to match the extents of the MFU. Therefore, we consider wavepackets of size $(L_x^+,L_y^+,L_z^+) = (320.5, 40, 160.2)$. As the FC domain is considerably larger than the MFU, we can track many wavepackets simultaneously in each flow field, as seen in figure \ref{fig:SchematicIdea}(b).

\subsection{Production--dissipation trajectories and advection speed}

We first consider trajectories of production and dissipation,
\begin{equation}
    \llangle P\rrangle = -\frac{1}{L_y}\int_{0}^{L_y}\langle uv\rangle \frac{\partial U}{\partial y}\,\mathrm{d}y, \quad \llangle \varepsilon\rrangle  = \frac{1}{L_y}\int_{0}^{L_y} 2\nu \langle S_{ij} S_{ij}\rangle\,\mathrm{d}y,
\end{equation}
in the minimal and full channels, where $S_{ij}$ is the rate-of-strain tensor, and $\langle\cdot\rangle$ indicates averages in the streamwise and spanwise directions in the domain of integration. 

\begin{figure}
     \centering
     \begin{subfigure}[b]{0.35\textwidth}
         \centering
         \includegraphics[width=\textwidth]{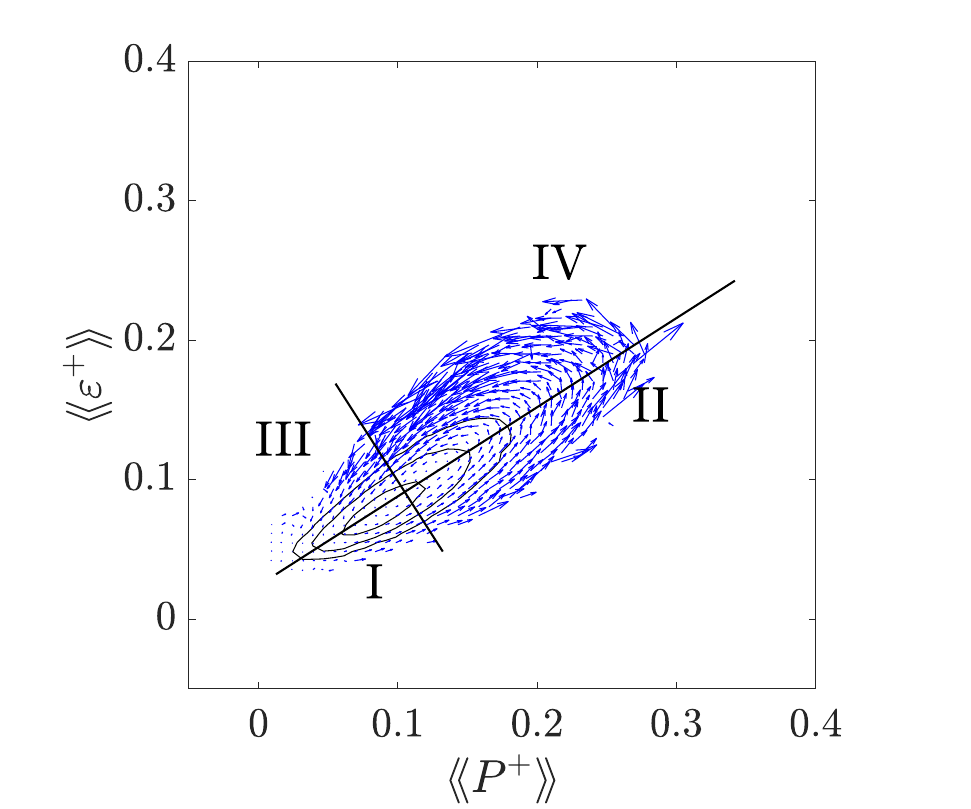}
         \caption{}
         \label{fig:PDstatistics_a}
     \end{subfigure}
     \begin{subfigure}[b]{0.35\textwidth}
         \centering
         \includegraphics[width=\textwidth]{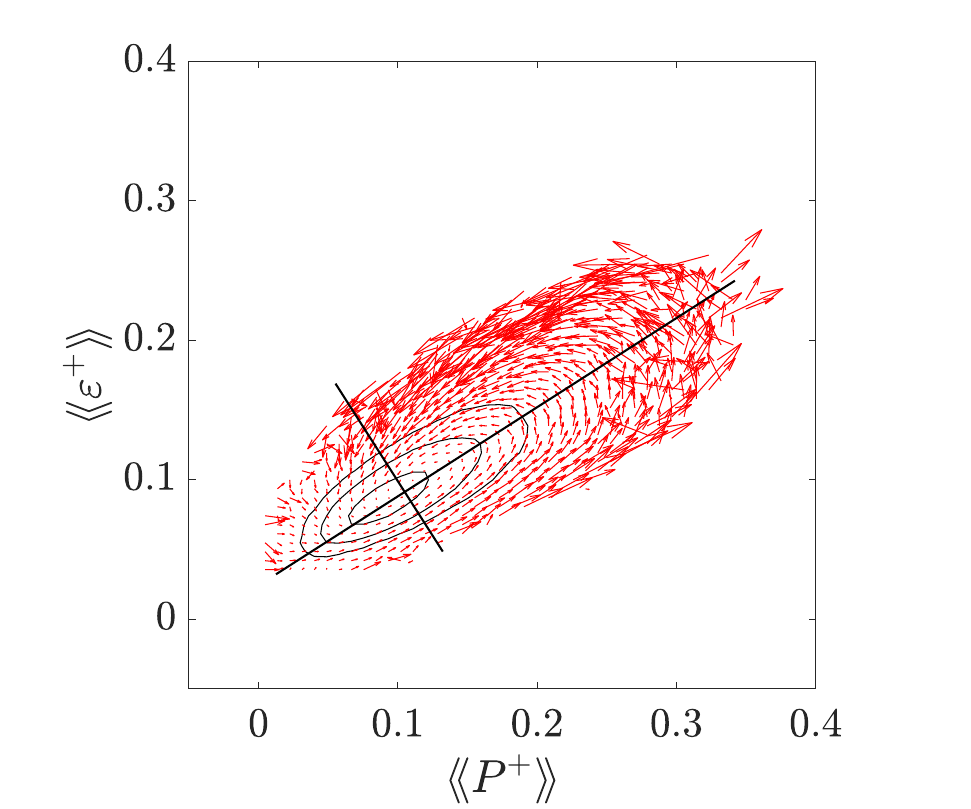}
         \caption{}
         \label{fig:PDstatistics_b}
     \end{subfigure}
     \begin{subfigure}[b]{0.35\textwidth}
         \centering
         \includegraphics[width=\textwidth]{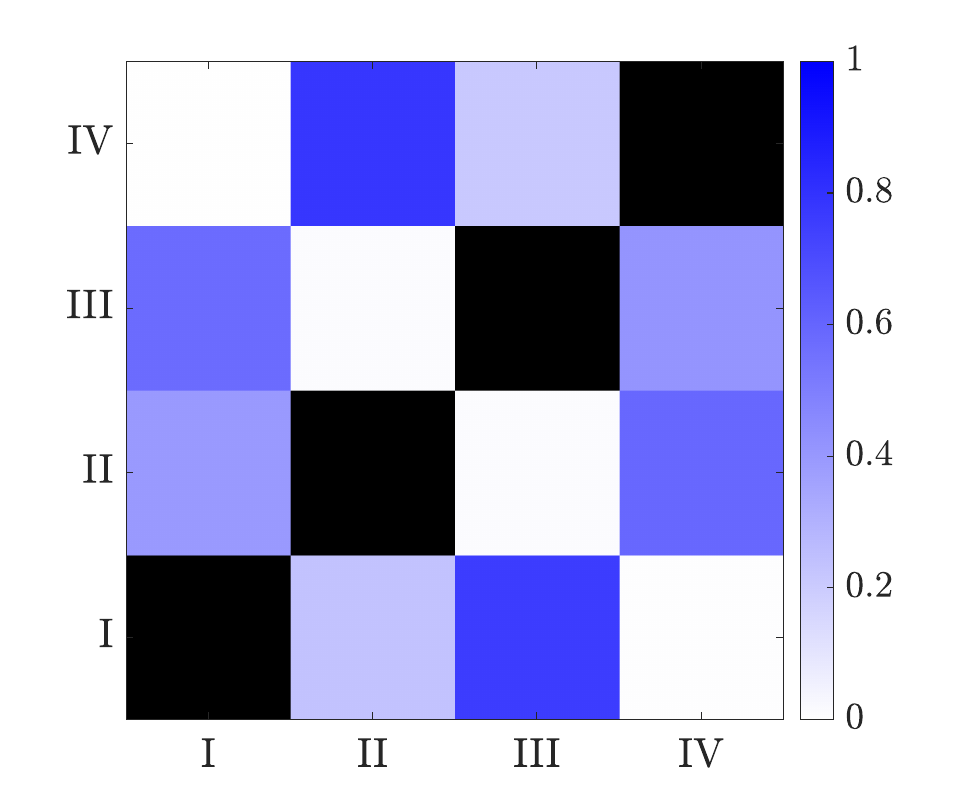}
         \caption{}
         \label{fig:PDstatistics_c}
     \end{subfigure}
     \begin{subfigure}[b]{0.35\textwidth}
         \centering
         \includegraphics[width=\textwidth]{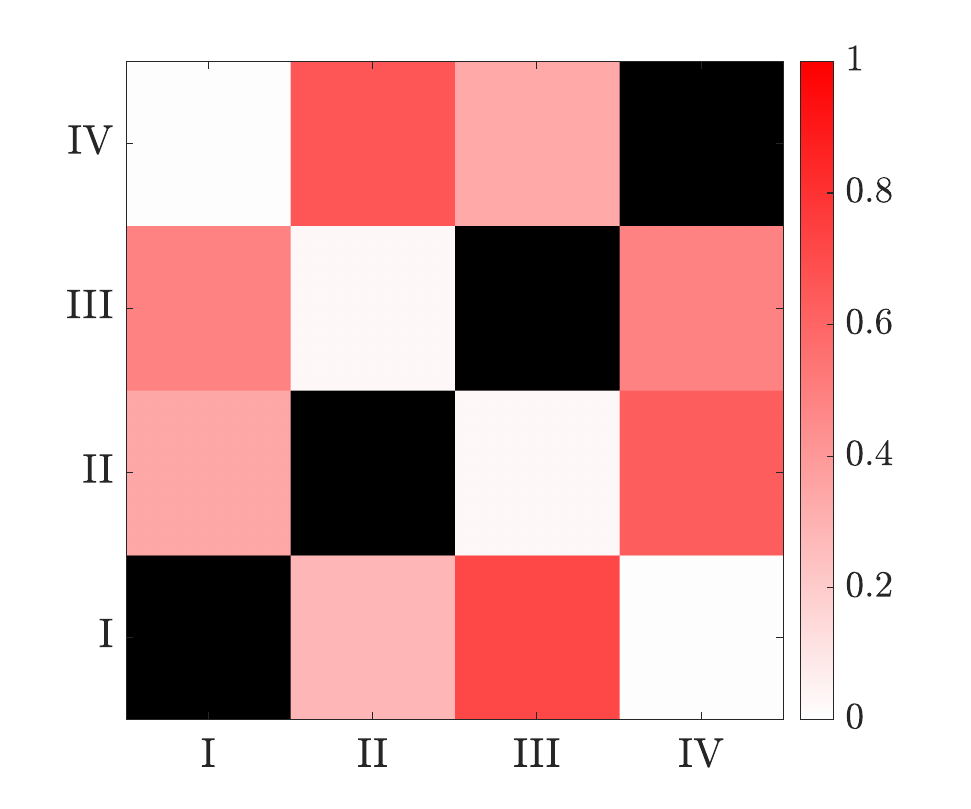}
         \caption{}
         \label{fig:PDstatistics_d}
     \end{subfigure}
        \caption{Production-dissipation of the (a) MFU and (b) FC. Contours are PDF at 10\%, 40\%, and 70\% of occurrences, and arrows show the mean trajectory computed by taking averages over $50\times 50$ bins. Perpendicular lines show quadrant divisions. (c) Relative transition probability between quadrants for the (c) MFU and (d) FC. Horizontal entries are the initial quadrant, and vertical entries are the next quadrant.}
        \label{fig:PDstatistics}
\end{figure}
The production and dissipation are calculated at each timestep for the MFU and the different instances of wavepackets in the FC. The joint PDF of those two quantities is shown in figure \ref{fig:PDstatistics}(a,b) for the two cases. The PDF is compiled as a histogram
over $50 \times 50$ bins, and the arrows represent the mean evolution velocity of all the states within a particular bin. These results are consistent with those observed in \cite{jimenez2005bursting}. We also observe that the mean evolution velocities of the MFU and FC are similar; however, the PDF of the FC has a larger spread of production and dissipation values than the MFU, indicating a much more intense production and dissipation events in the FC.

The production-dissipation space is then divided into quadrants following \cite{jimenez2005bursting}, with the origin at the mean production and dissipation, and the axes determined by the principal axes of the PDF. Although the FC has different mean production-dissipation values and principal axes compared to the MFU, the results were not sensitive to the choice of the mean values and principal axes, so the quadrants based on the MFU were used for consistency.

The probability of moving from one quadrant to another, referred to as the transition probability, was determined following the method of \cite{schmid2018} and is shown in figure \ref{fig:PDstatistics}(c,d). The transition probabilities only consider the relative probability of moving to different quadrants and do not consider the probability of staying in the same quadrant. Interestingly, figure \ref{fig:PDstatistics}(c,d) show a high probability of moving from $I$ to $III$ and from $IV$ to $II$, whereas (a,b) do not show arrows pointing in those directions. This is not necessarily a contradiction because figure \ref{fig:PDstatistics}(c,d) consider only the probability of moving from one quadrant to the other, while figure \ref{fig:PDstatistics}(a,b) also consider the speed of evolution of the state. To quantify the error between transition probabilities in the MFU and FC, the transition probabilities are assembled into a vector, and the error is defined as the norm of the difference in the probability vectors between the MFU and FC. 

It was noted that the choice of advection speed has a significant effect on the calculated quantities for the full channel. To determine the best advection method, we compare the production--dissipation plot for the MFU and FC using different advection methods and compare the transition probabilities.
The advection velocity that minimized error was the average velocity of only the regions of the flow contained in low-speed streaks, defined as 
\begin{equation}\label{eq:strkdefinition}
    \{(x,y,z): \sqrt{u^2 + w^2} > \alpha u_\tau \quad \& \quad u<0\}\mathrm{,}
\end{equation} 
at $y^+=40$, with $\alpha=3.4$ \cite{bae2021}. Other definitions of advection velocities were considered, including average velocity over the whole box, average velocity at several $y$-locations, and average velocity conditioned to negative $u$-fluctuations, but all had higher errors in the production-dissipation cycle. Interestingly, \cite{lozano2014time} shows that these structures advect with the mean velocity at $y^+=40$. However, in the case of production-dissipation cycles, this velocity resulted in higher error than the velocity conditioned to streaks at $y^+=40$. For the remainder of the section, we use the definition given by equation \eqref{eq:strkdefinition}. This leads to an average advection velocity in the streamwise direction of $10.7 u_\tau$ which is similar, but not exactly equal to, the advection velocity of slow streaks extracted in the FC using coherent structure tracking as presented in table \ref{tab:AdvectionVelocityLifeTime}. Note that the advection velocity reported in table \ref{tab:AdvectionVelocityLifeTime} is of the volumetric centroid of the streak, i.e., its group velocity, rather than the average of the velocity field occupied by the region of the slow streak only at $y^+=40$. 

\subsection{Temporal motif of energy contained in streak, roll, and small scales}

Additionally, we track energy in eddies related to the self-sustaining process. The magnitude of the streaks is defined by the magnitude of the streamwise velocity captured in the $(k_x, k_z) \equiv (k_x^*L_x/(2\pi), k_z^*L_z/(2\pi)) = (0,1)$ mode of the MFU domain,
\begin{equation}
    \llangle \hat{u}_{0,1} \rrangle = \frac{1}{L_y} \int_{0}^{L_y} (\hat{u}(k_x=0, k_z=1)^2)^{1/2}\,\mathrm{d}y,
\end{equation}
where $k_x^*$ and $k_z^*$ are the dimensional wavenumbers.
Similarly, the magnitude of the rolls is defined by the wall-normal and spanwise velocity magnitude of the $(k_x,k_z) = (0,1)$ mode of the MFU domain,
\begin{equation}
    \llangle \hat{r} \rrangle = \frac{1}{L_y} \int_{0}^{L_y} \left(\hat{v}(k_x=0, k_z=1)^2 + \hat{w}(k_x=0, k_z=1)^2\right)^{1/2}\,\mathrm{d}y.
\end{equation}
Finally, the magnitude of the small scales is defined as the sum of all velocity magnitude contained in smaller scales,
\begin{equation}
    \llangle \hat{s} \rrangle =\frac{1}{L_y} \int_{0}^{L_y} \left(\sum_{(k_x,k_z)\neq (0,0),(0,\pm 1)} \hat{u}(k_x,k_z)^2+\hat{v}(k_x,k_z)^2+\hat{w}(k_x,k_z)^2 \right)^{1/2} \,\mathrm{d}y.
\end{equation}
The definitions are consistent with the definition of a streak and roll in an MFU, where a single pair of infinitely long low- and high-speed streaks are present.

\begin{figure}
     \centering
     \begin{subfigure}[b]{0.35\textwidth}
         \centering
         \includegraphics[width=\textwidth]{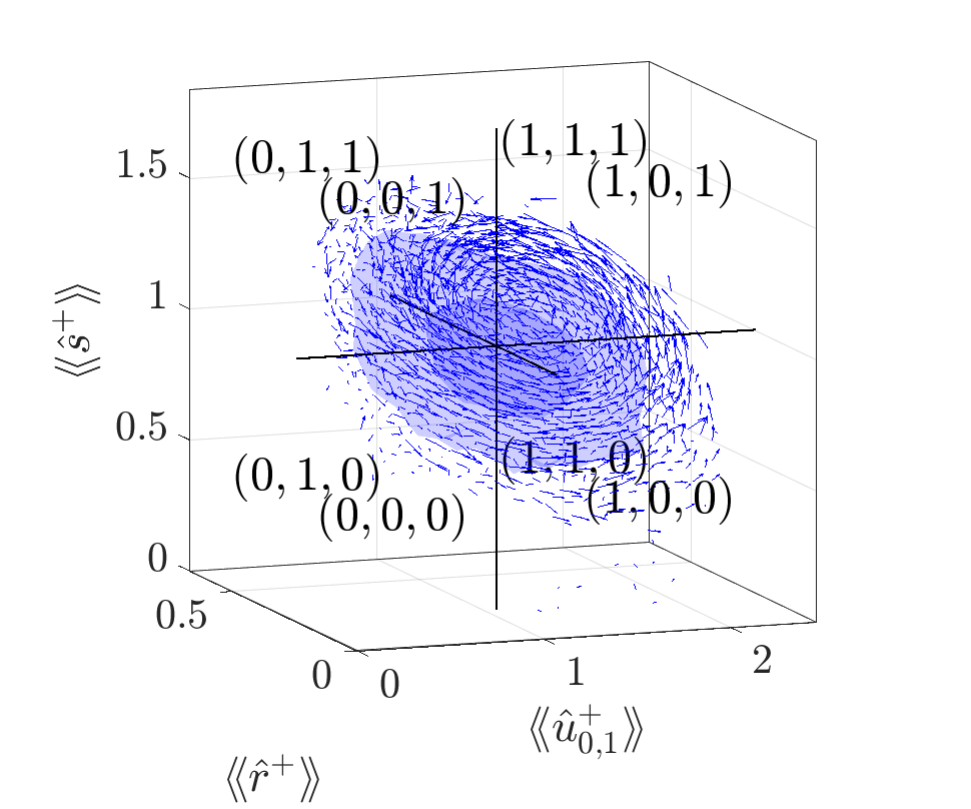}
         \caption{}
     \end{subfigure}     
     \begin{subfigure}[b]{0.35\textwidth}
         \centering
         \includegraphics[width=\textwidth]{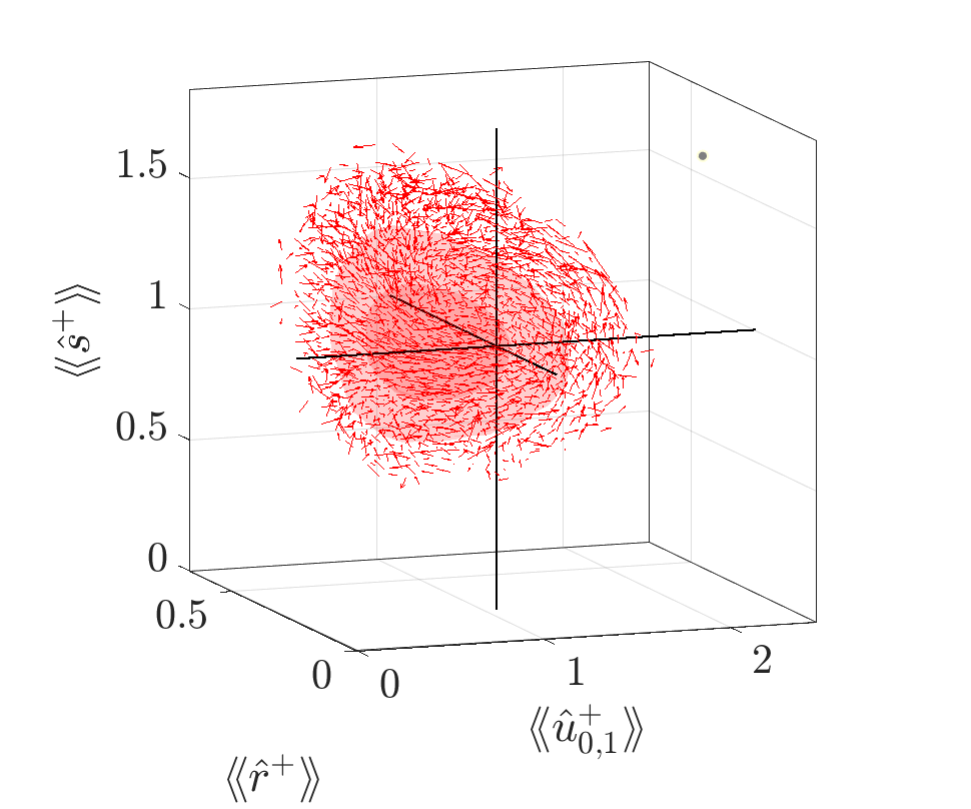}
         \caption{}
     \end{subfigure}
     \begin{subfigure}[b]{0.35\textwidth}
         \centering
         \includegraphics[width=\textwidth]{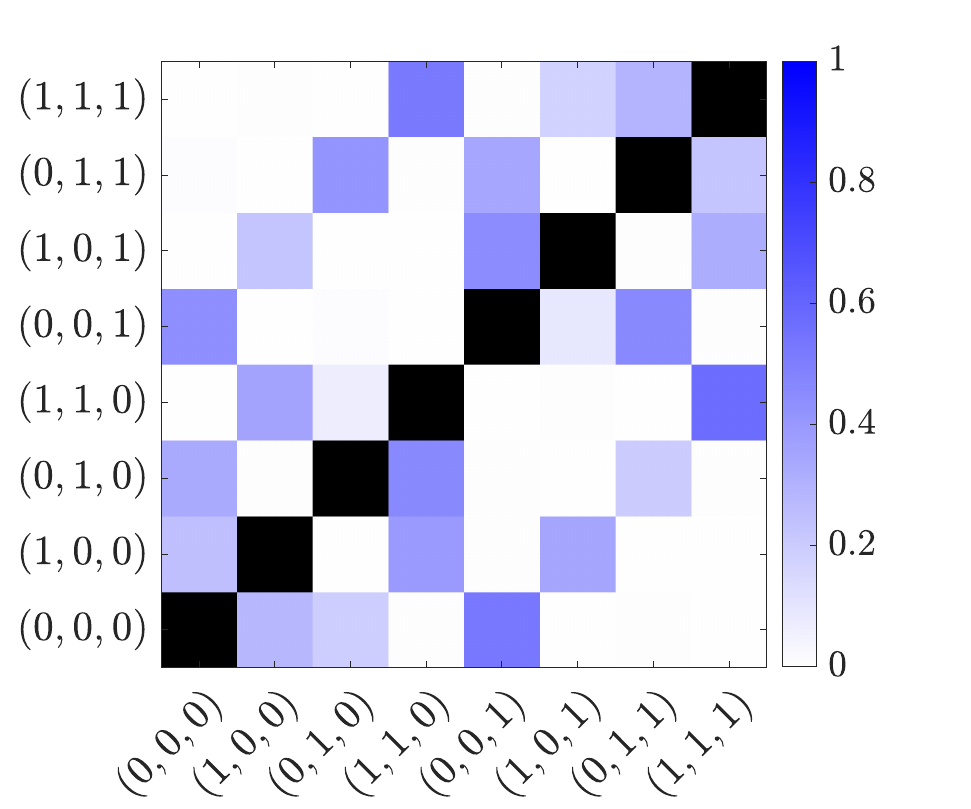}
         \caption{}\label{fig:Octantdiagram}
     \end{subfigure}
     \begin{subfigure}[b]{0.35\textwidth}
         \centering
         \includegraphics[width=\textwidth]{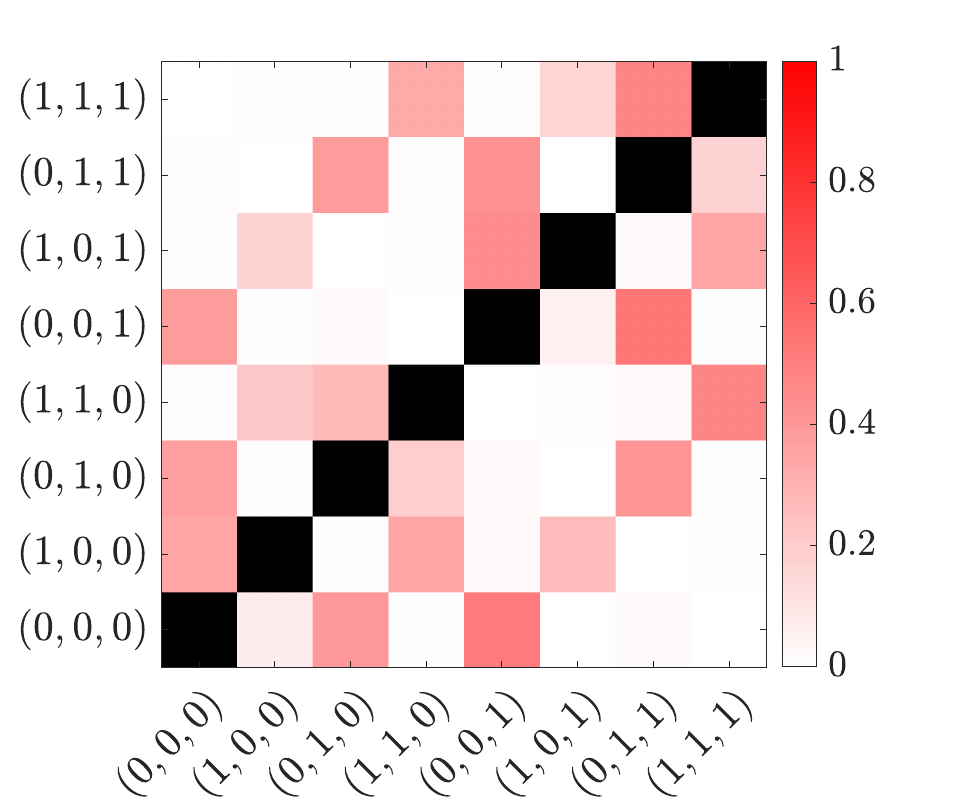}
         \caption{}
     \end{subfigure}
        \caption{Magnitude of the streak, roll, and small scales for the (a) MFU and (b) FC. Contours are at 10\% and 50\% of occurrences, and arrows show the mean trajectory. Black lines show octant divisions. Relative transition probability between octants for the (c) MFU and (d) FC. Horizontal entries are the initial octant, and vertical entries are the next octant.}
        \label{fig:Octantstatistics}
\end{figure}

The magnitude of streaks, rolls, and small scales are analyzed similarly to production and dissipation. Distribution of the quantities and average trajectories are shown in figure \ref{fig:Octantstatistics}(a,b). Since three quantities were considered, the phase space was divided into eight octants, with the origin at the mean value of all three for the MFU. Principal axes of the data were identified, but they were very closely aligned with the Cartesian axes, so Cartesian axes were used to divide the space into octants, as shown in figure \ref{fig:Octantstatistics}(a,b). The octants were named with a 3-digit binary key, with 0 for below-average magnitude and 1 for above-average magnitude. The order of entries is (streaks, rolls, small scales), so for example, octant $(0,1,1)$ represents low magnitude contained in streaks, high magnitude contained in rolls, and high magnitude contained in small scales. Transition probabilities were calculated between the octants, shown in \ref{fig:Octantstatistics}(c,d). Generally, transition probabilities in the MFU and FC have many similarities. The MFU has a larger spread of probability values, and it has more asymmetry in the direction of movement between octants. For example, $(0,1,0)\to(0,0,0)$ has a significantly higher probability than $(0,0,0)\to(0,1,0)$, but in the FC, both directions have a similar probability.

\begin{figure}
     \centering
     \begin{subfigure}[b]{0.32\textwidth}
         \centering
         \includegraphics[width=\textwidth]{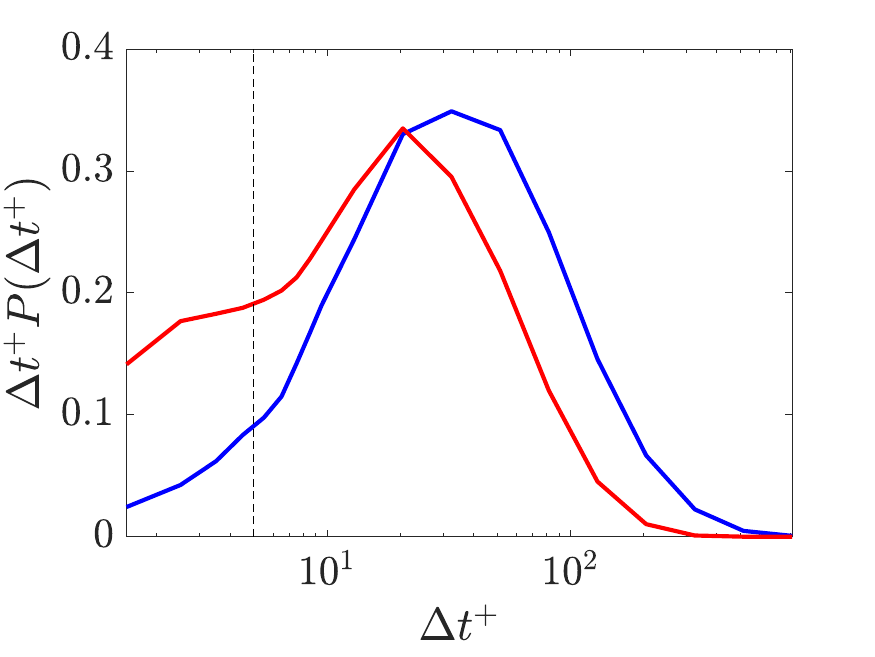}
         \caption{}
     \end{subfigure}
     \begin{subfigure}[b]{0.32\textwidth}
         \centering
         \includegraphics[width=\textwidth]{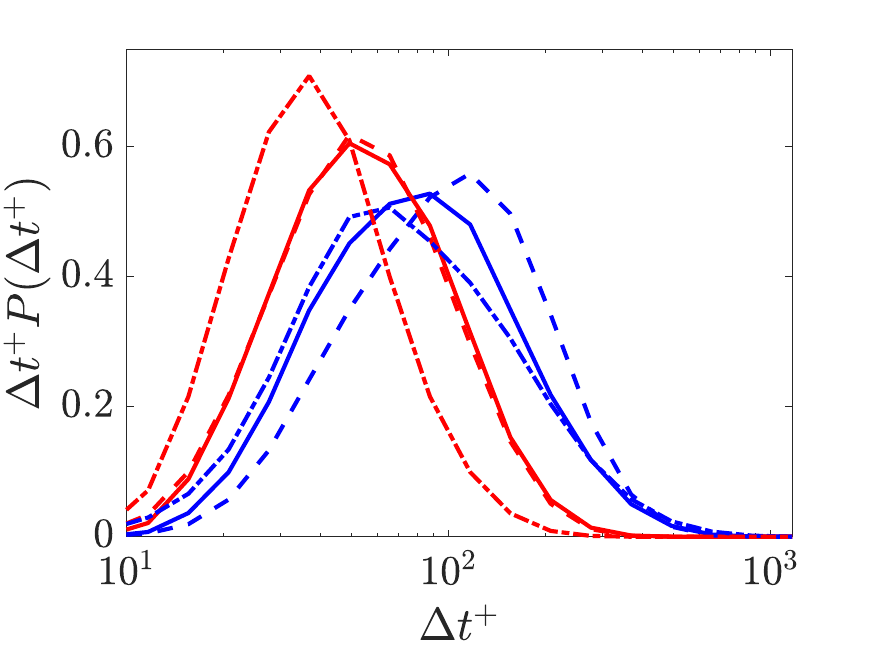}
         \caption{}
     \end{subfigure}
     \begin{subfigure}[b]{0.32\textwidth}
         \centering
         \includegraphics[width=\textwidth]{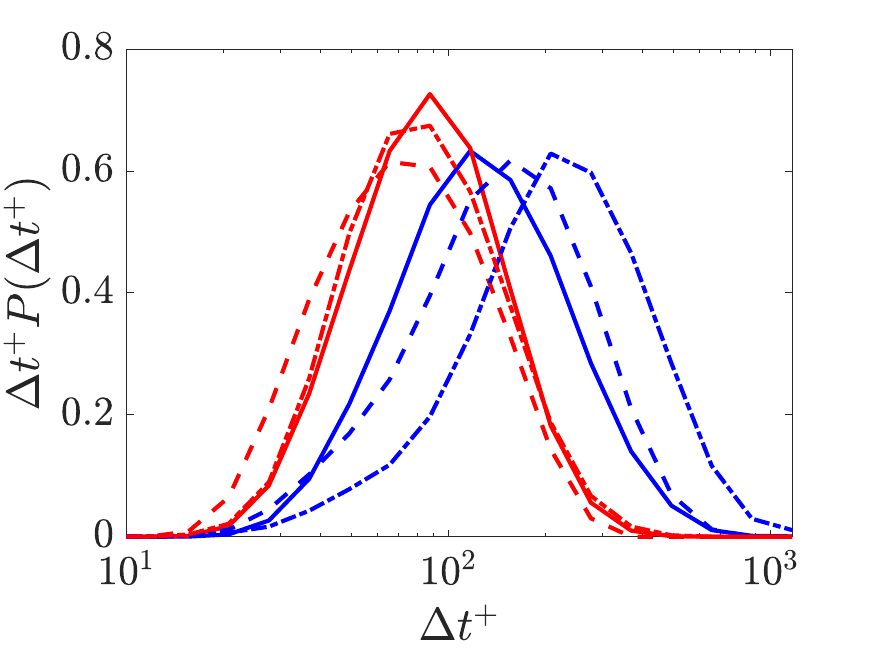}
         \caption{}
     \end{subfigure}
        \caption{(a) PDF of duration of single events for the MFU (blue) and FC (red). The vertical line at $\Delta t^+=5$ indicates the filter cutoff. (b) PDF of duration of two-event subgraphs for the MFU (blue) and FC (red). Solid lines represent all two-event subgraphs, dashed lines represent the $(1,1,0)\to(1,1,1)$ motif, and dot-dashed lines represent the $(1,0,0)\to(1,0,1)$ motif. (c) PDF of duration of three-node subgraphs for the MFU (blue) and FC (red). Solid lines represent all three-event subgraphs, dashed lines represent the $(0,0,1)\to(1,0,1)\to(1,0,0)$ motif, and dot-dashed lines represent the $(0,0,0)\to(1,0,0)\to(1,1,0)$ motif.}
        \label{fig:OctantTiming}
\end{figure}

We then form a graph using the trajectories. Each time the trajectory enters a new octant, it is defined as an event. The PDF of the duration of single events is shown in \ref{fig:OctantTiming}(a). Events in the MFU had longer durations than the FC, with the average duration of an MFU event being 2.21 times as long as the average FC event. This value is close to the ratio of lifetimes of the spatiotemporal motifs reported in table \ref{tab:AdvectionVelocityLifeTime} of $2.83$ and is even closer to the ratio of timescales implied by the advection velocities of the slow streaks between the two flows of approximately $2.17$. The trajectory can be recast as a series of events, which then form the graph, where each event is a node and the transition from one event to the next event is an edge between the two nodes. Some events occurred for short durations, especially in the FC, so events lasting $\Delta t^+ \le 5$ were removed, and the trajectory was reassembled. This corresponds to the region of figure \ref{fig:OctantTiming}(a) where the duration of events in the FC started to plateau. It was found that the motif results were not sensitive to a cutoff value of $\Delta t^+ = 10$. The resulting graph sizes were 17,952 events for the MFU and 90,579 events for the FC.

The graphs were analyzed for motifs following the method of \cite{Milo2002}. Because each node in the graph had one incoming and one outgoing connection, the method could be simplified and was much less computationally expensive than the method described in section \ref{CST}.  Due to the restricted nature of the trajectories, there were $8^n$ possible isomorphic subgraphs of $n$ nodes. To analyze the trajectory for $n$-node motifs, the trajectory was broken down into sequences of $n$ event-induced subgraphs, each classified as one of the $8^n$ isomorphic subgraphs. The number of times each isomorphic subgraph occurred was then tabulated. The significance of each of the induced subgraphs were computed following equation \ref{SignificanceScore}.

The first five 2-node and 3-node subgraphs with the highest significance are reported in tables \ref{tab:2eventmotifs} and \ref{tab:3eventmotifs}. Generally, motifs in the MFU had a higher significance $\widetilde{W}(s)$ compared to the FC. This was expected since the MFU is more dynamically restricted, so the dynamics were frequently repeated.

When considering 2-node motifs, all identified motifs in the MFU, and three of the top five motifs in the FC, involve changing energy in the small scales. subgraphs $(1,1,0) \to (1,1,1)$ and $(1,0,0) \to (1,0,1)$ were among the five most significant motifs in both the MFU and FC. These motifs represent physical phenomena, with the flow initially containing high energy in large-scale structures and moving to a state with energy in both large and small-scale structures, reflective of the energy cascade.

The 2-node motifs provide insight into the directionality of the trajectories. For example, in the MFU, $(0,0,0) \to (0,0,1)$ and $(0,0,1) \to (0,0,0)$ have very similar significance values, which suggests that both directions have similar importance to the dynamics. However, $(0,1,1)\to(0,1,0)$ has much higher significance than $(0,1,0)\to(0,1,1)$, with significance scores $\widetilde{W}_{(0,1,1)\to(0,1,0)}=0.3276$ and $\widetilde{W}_{(0,1,0)\to(0,1,1)}=0.0236$. This matches the physical understanding of the SSP because $(0,1,1)\to(0,1,0)$ reflects the step in the SSP where streamwise rolls are formed from small-scale structures by a nonlinear process. $(0,1,0)\to(0,1,1)$ is not understood as a part of the SSP, so it has a lower significance. Interestingly, there are variations in directionality between the MFU and FC. For example, in the FC, $(1,0,0)\to(1,0,1)$ and $(1,0,1)\to(1,0,0)$ are similar, with 
\begin{equation*}
    \frac{\widetilde{W}_{(1,0,0)\to(1,0,1)}}{\widetilde{W}_{(1,0,1)\to(1,0,0)}}=1.95.
\end{equation*}
In the MFU, $(1,0,0)\to(1,0,1)$ is much more significant than $(1,0,1)\to(1,0,0)$, with \begin{equation*}
    \frac{\widetilde{W}_{(1,1,0)\to(1,1,1)}}{\widetilde{W}_{(1,1,1)\to(1,1,0)}}=4.54.
\end{equation*}
When considering the ratios of the number of occurrences, rather than the significance, similar values are found. This suggests that the directionality of the movement between these octants is more significant in the MFU than the FC.

When considering 3-node motifs, several very rare subgraphs had high significance because they were also very rare in the randomly generated networks. Therefore, we did not report subgraphs that occurred fewer than 40 times. These rare subgraphs tended to contain diagonal movements or back-and-forth movements between octants. The five subgraphs with the highest significance meeting the occurrence threshold are reported in table \ref{tab:3eventmotifs}. Again, there is strong similarity between the MFU and FC, and the MFU tended to have higher significance values.

Motifs $(0,0,1)\to(1,0,1)\to(1,0,0)$ and $(0,0,0)\to(1,0,0)\to(1,1,0)$ were among the five most significant motifs in both the MFU and FC. While most subgraphs had similar significance rankings between the MFU and FC, there were a few that were very different. For example, $(0,0,0)\to(0,0,1)\to(1,0,1)$ had the third highest significance in the MFU, with $\widetilde{W}_{(0,0,0)\to(0,0,1)\to(1,0,1)}=0.0410$.
However in the FC, $(0,0,0)\to(0,0,1)\to(1,0,1)$ was the 349th most significant subgraph, with a negative significance value of $\widetilde{W}_{(0,0,0)\to(0,0,1)\to(1,0,1)}=-.0013$. This suggests that there are dynamical patterns that are very significant to the MFU but not the FC. This is similar to the re-ordering of the most significant, as well as most common, yet dominant, spatiotemporal motifs in table \ref{tab:SpatioTempMotifZFreq}. However, in general, the similarities between the motifs indicate that there is significant dynamical similarity between the MFU and FC.

The duration of significant motifs was compared to the duration of all subgraphs taken together, and the results are shown in figure \ref{fig:OctantTiming}(b,c). Duration was determined by summing the duration of the events making up the subgraph. In particular, motifs that were among the top five most significant in both the MFU and FC were considered. For two node motifs, this was $(1,1,0)\to(1,1,1)$ and $(1,0,0)\to(1,0,1)$, and for three node motifs, this was $(0,0,1)\to(1,0,1)\to(1,0,0)$ and $(0,0,0)\to(1,0,0)\to(1,1,0)$. The FC tended to have faster times than the MFU, and this was seen in both the duration of all events, and the duration of individual subgraphs. Even though the same subgraphs were occurring with similar significance values in the MFU and FC, they tended to have shorter durations in the FC. Further, the durations of the significant subgraphs did not scale proportionally between the MFU and FC. For example, $(0,0,1)\to(1,0,1)\to(1,0,0)$ was generally slower than the background turbulence (represented by the solid lines in figure \ref{fig:OctantTiming}) in the FC, but faster than the background turbulence in the MFU. 

\begin{table}
\caption{\label{tab:2eventmotifs} Five most significant 2-node motifs. Bold-face motifs are common between the MFU and FC. $\widetilde{W}(s)$ is the significance score, and Occ., or occurrence, is the number of times the subgraph occurs in the real trajectory.}
\begin{center}
\renewcommand{\arraystretch}{1} 
\begin{tabular}{clrlclr}
\br
\multicolumn{3}{c}{Minimal Flow Unit}                                                                      && \multicolumn{3}{c}{Full Channel}                                                                          \\ \mr
$s$                                                                                       & $\widetilde{W}(s)$ & Occ. && $s$                                                                                       & $\widetilde{W}(s)$ & Occ. \\ \mr
$\boldsymbol{(1, 1, 0) \rightarrow (1, 1, 1)}$                                            & 0.370  & 1669 && \begin{tabular}[c]{@{}c@{}}$\boldsymbol{(1, 1, 0) \rightarrow (1, 1, 1)}$\end{tabular} & 0.326  & 2664 \\
$(0,1,1)\rightarrow(0,1,0)$                                                               & 0.328  & 816 && ${(1, 1, 0) \rightarrow (1, 0, 0)}$ & 0.244  & 1181 \\
${(0, 0, 0) \rightarrow (0, 0, 1)}$ & 0.297  & 1109  && $(1,1,1)\rightarrow(1,1,0)$                                                               & 0.229  & 2024 \\
$(0,0,1)\rightarrow(0,1,1)$                                                               & 0.290  & 1141 && \begin{tabular}[c]{@{}c@{}}$\boldsymbol{(1, 0, 0) \rightarrow (1, 0, 1)}$\end{tabular}                                                               & 0.191  & 526 \\
\begin{tabular}[c]{@{}c@{}}$\boldsymbol{(1, 0, 0) \rightarrow (1, 0, 1)}$\end{tabular}                                                               & 0.279  & 610 && ${(1, 1, 1) \rightarrow (1, 0, 1)}$& 0.178  & 888 \\ \br
\end{tabular}
\end{center}
\end{table}

\begin{table}[h]
\caption{\label{tab:3eventmotifs}Five most significant 3-node motifs, excluding those with fewer than 12 occurrences. Bold-face motifs are common between the MFU and FC. $\widetilde{W}(s)$ is the significance score, and Occ., or occurrence, is the number of times the subgraph occurs in the real trajectory.}
\begin{center}
\renewcommand{\arraystretch}{1} 
\begin{tabular}{clrlclr}
 \br
 \multicolumn{3}{c}{Minimal Flow Unit}                                                                      && \multicolumn{3}{c}{Full Channel}          \\
 \mr
  $s$ & $\widetilde{W}(s)$ & Occ. && $s$ &  $\widetilde{W}(s)$ & Occ.\\
  \mr  
\small$\mathbf{(0,0,1)\to(1,0,1)\to(1,0,0)}$&0.0466&79&&\small${(0,1,1)\to (1,1,1)\to(1,1,0)}$&0.0178&910\\
\small${(0,1,0)\to(1,1,0)\to(1,0,0)}$&0.0432&283&&\small$\mathbf{(0,0,0)\to (1,0,0)\to(1,1,0)}$&0.0167&282\\
\small$(0,0,0)\to(0,0,1)\to(1,0,1)$&0.0410&138&&\small${(1,0,1)\to(1,1,1)\to(1,0,1)}$&0.0156&127\\
\small$(1,1,1)\to(1,1,0)\to(0,1,0)$&0.0305&89&&\small$\mathbf{(0,0,1)\to(1,0,1)\to(1,0,0)}$&0.0153&106\\
\small$\mathbf{(0,0,0)\to(1,0,0)\to(1,1,0)}$&0.0296&261&&\small$(0,0,1)\to(1,0,1)\to(1,1,1)$&0.0132&206\\
 \br
\end{tabular}
\end{center}
\end{table}

\section{Conclusions}\label{Conclusions}

Interpreting wall-bounded turbulence as a collection of interacting coherent structures is a long-standing paradigm. However, it needs a consistent framework or an algorithmic approach to extract and analyze important dynamical events. In this work, we attempt to do so by employing coherent-structure tracking, semi-Lagrangian wavepackets, and analysis techniques from network science and graph theory. As a proof of concept, we had two aims. First, to extract the signature of the self-sustaining process algorithmically. Second, to analyze the dynamics of the extracted self-sustaining process in both a minimal flow unit and a large unrestricted domain to understand how they differ. 

To do so, we first extended the coherent structure tracking algorithms of \cite{lozano2014time} to allow for simultaneously tracking multiple sets of eddies. In this study, we limited ourselves to tracking attached ejections, sweeps, and slow and fast streaks simultaneously. The spatiotemporal interactions between these eddies are encoded into a complex multilayer network, which is then searched for significantly non-random patterns given an appropriate model for a random network, referred to as network motifs \cite{Milo2002}. This analysis extracted the roll-streak pairing as the most significant and perhaps simplest non-random dynamical pattern in both flows, satisfying the first goal. However, it was detected that other significant, as well as frequent, yet dominant patterns were ordered differently between the two flows. This indicates that while the most prominent dynamical mechanism is the same between the MFU and the FC, there are others that are substantially different in both frequency and significance between the two flows. Furthermore, even for the most significant dynamical pattern, $\mathcal{S}_1^{\{F,M\}}$, the average timescale was $2.83$ longer in the MFU than the FC. 

We simultaneously employed semi-Lagrangian wavepackets to analyze the phase-space dynamics of entire MFU-sized regions advecting within the FC, and compared them to those of the MFU. Within the octant space defined by streak, roll, and small-scale magnitude, graphs were constructed from the sequence of transition events from one octant to the next. Overall, it was found that both the 2-node and 3-node motifs were mostly similar between the MFU and the FC, with 3-node motifs representing transitions from one scale to another. However, a few key differences were found. First, the significance scores, $\widetilde{W}(s)$, were higher for the MFU compared to FC, presumably due to the restricted dynamics of the MFU. Second, the average timescales of each motif were once again faster in the FC compared to MFU, with the average duration of an MFU event being $2.21$ slower than that of an FC event, consistent with the motif lifetimes extracted using spatiotemporal tracking, and in particular with the implied ratio of timescales for the slow speed streaks of $2.17$. 

Overall, an algorithmic multi-eddy-type spatiotemporal tracking approach to detecting key dynamical mechanisms in wall-bounded turbulence was proposed, developed, and tested on a control experiment. It was found that the roll-streak SSP is the lowest form of spatiotemporal organization, and that its timescale in the MFU is substantially slower compared to its unrestricted counterpart. This key finding was corroborated with a semi-Lagrangian wavepacket approach combined with temporal motif extraction in phase space. In the future, we aim to extend this approach to detect key interscale dynamics at higher $Re_\tau$, and extract higher-order mechanisms by comparing to more complex random network models that constrain more than just the colored-degree distribution.

\section*{Acknowledgments}

A. Elnahhas and P. Moin acknowledge support from NASA's Transformational Tools and Technologies project under grant no. 80NSSC20M0201. E. Lenz acknowledges support from the NSF Graduate Research Fellowship under grant no. DGE-1745301. All authors acknowledge insightful comments from Jos{\'e} Cardesa on an earlier version of the manuscript. All authors acknowledge the European Research Council (ERC) under the Caust grant ERC-AdG-101018287, and Prof. J. Jim{\'e}nez for the participation in the 2023 Madrid Summer Turbulence Workshop.

\end{document}